\documentclass{AIAA}
\usepackage{hyperref}

\newcommand\diff{\mathrm{d}}

\newcommand\Pran{\mbox{\textit{Pr}}} % Prandtl number, cf TeX's \Pr product
\newcommand{\RN}[1]{%
	  \textup{\uppercase\expandafter{\romannumeral#1}}%
  }

\begin{document}

\title[DNS of transitional and turbulent SBLI]{Numerical investigation of supersonic shock-wave/boundary-layer interaction in transitional and turbulent regime}

\author{R. Quadros\footnote{Research Fellow, Dipartimento di Ingegneria Meccanica e Aerospaziale, russell.quadros@uniroma1.it} and
        M. Bernardini\footnote{Assistant Professor, Dipartimento di Ingegneria Meccanica e Aerospaziale, matteo.bernardini@uniroma1.it}}
\affiliation{Dipartimento di Ingegneria Meccanica e Aerospaziale, Via Eudossiana 18, 00184, Roma, Italia.}

\begin{abstract}

We perform direct numerical simulations of shock-wave/boundary-layer interactions (SBLI)
at Mach number $M_{\infty} = 1.7$ to investigate the influence of the state of the incoming boundary layer
on the interaction properties. We reproduce and extend the flow conditions of the experiments performed
by \citet{giepman16}, in which a spatially evolving laminar boundary layer over a flat plate is initially tripped
by an array of distributed roughness elements and impinged further downstream by an oblique shock wave.
Four SBLI cases are considered, based on two different shock impingement locations
along the streamwise direction, corresponding to transitional and turbulent interactions,
and two different shock strengths, corresponding to flow deflection angles $\phi=3^o$ and $\phi=6^o$.
We find that, for all flow cases, shock induced separation is not observed, the boundary layer remains attached
at $\phi = 3^{\circ}$ and close to incipient separation at $\phi = 6^{\circ}$, independent of the state
of the incoming boundary layer. We characterize the regions of instantaneous separation by
computing the statistical probability ($\overline{\gamma}$) of the wall points with local flow reversal.
The analysis shows that the turbulent interactions are characterized by a higher peak of $\overline{\gamma}$,
although the region of separation is slightly wider in the transitional interaction cases.
The extent of the interaction zone is mainly determined by the strength of the shock wave, and the state of the incoming
boundary layer has little influence on the interaction length scale $L$. The scaling analysis for $L$ and the separation criterion
developed by \citet{souverein13} for turbulent interactions are found to be equally applicable for the transitional interactions.
The findings of this work suggest that a transitional interaction %retains most of the beneficial features of a turbulent SBLI and it
might be the optimal solution for practical SBLI applications, as it removes the large separation bubble typical of laminar
interactions and reduces the extent of the high-friction region associated with an incoming turbulent boundary layer. 

\end{abstract}

%\keywords{}


\maketitle
\section*{Nomenclature}
%(Nomenclature entries should have the units identified)\\
\noindent\begin{tabular}{@{}lcl@{}}

$M_{\infty}$  &=& freestream Mach number \\
$c_f$  &=& skin friction coefficient  \\
$c_p$  &=& specific heat at constant pressure   \\
$k_c$  &=& thermal conductivity  \\
$G_3$  &=& scaling parameter for the interaction lengthscale\\
$k$    &=& height of roughness elements  \\
%$\overline{k}$  &=& constant used in scaling analysis  \\
$L$    &=& interaction length scale  \\
$L_x$  &=& domain length in streamwise direction \\
$L_y$  &=& domain length in wall-normal direction\\
$L_x,L_y,L_z$  &=& size of the computational domain in the streamwise, wall-normal and spanwise directions \\
$P$  &=& mean pressure   \\
$Pr$  &=& molecular Prandtl number   \\
$p_{w_{rms}}$  &=& rms value of wall pressure fluctuation \\
$q_{\infty}$  &=& freestream dynamic pressure   \\
$Re_\theta$  &=& Reynolds number based on momentum thickness  \\
$Re_{x_{t}}$  &=& Reynolds number based on the inlet-trip distance   \\
$R$  &=& specific gas constant   \\
$S_e^*$  &=& separation criterion \\
$T$  &=& mean temperature  \\
$U$  &=& mean velocity   \\
$x,y,z$  &=& Cartesian coordinates in the streamwise, wall-normal and spanwise directions \\
$x_s$  &=& distance from the domain inlet to the shock impingement location \\
$x_t$  &=& distance from the domain inlet to the trip location  \\
$\beta$  &=& shock angle   \\
$\delta_v$  &=& viscous length scale  \\
$\delta_{in}$  &=& inlet boundary-layer thickness \\
$\delta_{95}$  &=& boundary-layer thickness based on 95 \% of the external velocity  \\
$\delta^*_i$  &=&  incompressible displacement thickness  \\
$\theta_i$  &=& incompressible momentum thickness \\
$H_i$  &=& incompressible shape factor  \\
%$\Delta$  &=&  wall units \\
$\gamma$  &=& specific heat ratio  \\
$\mu$  &=& molecular viscosity \\
$\overline{\gamma} $  &=& statistical probability of reverse flow\\
$\phi$  &=& flow deflection angle \\
$\rho$  &=& density \\
$\tau$  &=& wall shear stress \\
\end{tabular} \\

\subsection*{subscripts}
\noindent\begin{tabular}{@{}lcl@{}}
$\infty$  &=& evaluated in the freestream  \\
$k$  &=& evaluated at roughness edge  \\
$w$  &=& evaluated at wall \\
\end{tabular} \\

\subsection*{superscripts}
\noindent\begin{tabular}{@{}lcl@{}}
$+$  &=& normalization in wall units  \\
\end{tabular} \\

\section{Introduction}
\label{sec_intro}

Shock-wave/boundary-layer interactions (SBLI) are commonly observed in high speed engineering applications
such as air intakes, turbo-machinery cascades, helicopter blades, supersonic nozzles, and launch vehicles.
Shock waves can be useful for compressing the incoming flow, enhancing the turbulence mixing and increasing
the internal energy of the flow. However, their interaction with an incoming boundary layer could result in
boundary-layer separation, high-wall heat flux and surface pressure, and induction of large scale
instabilities~\citep{dolling01,dupont06,touber09,souverein_10_b,clemens14}. SBLI is also responsible for unsteady vortex shedding
and shock/vortex interaction, which are the major reasons for broadband noise emission.

The nature of the incoming boundary layer which interacts with the shock has a significant impact on the flow
topology and on the aerodynamic performance of the aerospace vehicle.
A laminar boundary layer is more susceptible to separation when it encounters a shock wave because of its
low resistance to the adverse pressure gradient created by the impinging shock \citep{adamson_messiter80,delery85}.
Such a boundary-layer separation could be a greater challenge in hypersonic intakes, where strong interactions occur leading to a
reduced mass flow rate \citep{babinsky09}. 
This laminar interaction has been studied in detail through experiments~\citep{hakkinen59,giepman15}, numerical simulations~\citep{degrez87,katzer89,yao07} and theory~\citep{gadd56}.
\citet{katzer89} showed that the separation size has a linear growth with the impinging shock strength and that, in agreement with the free interaction concept \citep{chapmanetal_57},
at low Reynolds numbers the separation size increases with Reynolds number and decreases with Mach number.

A possible strategy to reduce the separation size is to energize the incoming boundary layer to make it turbulent.
A turbulent boundary layer better sustains the adverse pressure gradient created by shock impingement, minimizing and
in some cases suppressing the flow separation \citep{greene70,stollery75,delery86}. 
Various techniques have been used in the past to energize the incoming boundary layer, which include suction and injection \citep{gai77,krogmann85}, slots \citep{holden05, smith04, raghu87} and vortex generators \citep{mccormick93, anderson06}.
Although a turbulent boundary layer could reduce the shock-induced separation size, an early onset of turbulence has the disadvantage of increasing the skin friction and the associated drag.

In recent years, there has been a renewed interest in studying the interaction of a transitional boundary layer with a shock wave, motivated by
the hope that a transitional interaction could bridge the gap between the large separation size obtained in a laminar interaction and the high friction
drag associated with a turbulent interaction. Some of the recent studies that have experimentally investigated the interaction of
a shock with a transitional boundary layer include those by \citet{sandham14,davidson14,davidson15,giepman15}.
~\citet{giepman15} studied the influence of the boundary-layer state (laminar, transitional and turbulent) on the properties of the interaction with an oblique impinging shock wave. 
The incoming laminar boundary layer transitioned to turbulence due to acoustic disturbances in the flow, and the oblique shock was impinged at varying locations based on the type of interaction desired.
The shock-impingement point in the transitional interaction case was selected at a location with intermittency of about 50 \%, and it resulted in a small boundary-layer separation. 
In a follow-up work, \citet{giepman16} used tripping devices to promote the boundary-layer transition ahead of the interaction.
They analyzed the effectiveness  of three tripping devices namely stepwise trip, zigzag strip and a patch of distributed roughness. 
The zigzag strip and the distributed-roughness patch were found to be more effective in energizing the boundary layer and suppressing the separation region, mainly due to their three-dimensional shape.

There has been a limited number of numerical studies on transitional SBLI in the literature, 
and the only prominent one pertains to the hypersonic regime.
\citet{sandham14} carried out direct numerical simulations of an oblique shock impinging on  
a transitional boundary layer at $M_\infty=6$, and compared the numerical results with 
available experimental data.
They triggered the boundary-layer transition by adding disturbances to the density field at the domain inlet.
%ATTENTION!!
The shock was impinged at various locations for varying intermittencies,
%and highest levels of wall heat transfer
%were consistently obtained for transitional rather than full
and they observed higher values of wall 
heat transfer for the transitional interactions as compared to the fully turbulent cases.
The objective of the present work is to numerically investigate the  flow physics of transitional supersonic shock/boundary-layer interaction.
In our simulations we use hemispherical roughness elements to trip an incoming laminar boundary layer which is impinged by an oblique shock at varying locations corresponding to
transitional or turbulent conditions. The flow configuration is chosen to match some of the experiments of~\citet{giepman16}.
We also analyze the effect of varying the shock strength on the transitional interaction  by increasing the shock generator angle.
The main aim is to highlight the advantages (if any) of a transitional interaction on the suppression of mean and instantaneous shock-induced separation,
and also to bring out the key properties of the interaction in terms of boundary-layer growth and interaction length scales.   

The paper is organized as follows. The numerical set up and the methodology are described in section~\ref{sec:methodology}.
Results pertaining to the case of the tripped boundary layer without any impinging shock are presented in section~\ref{sec:bl} along with a comparison with available experimental data.
The effect of shock impingement on the transitional and turbulent boundary layers is described in section~\ref{sec:sh}. 
Conclusions are finally provided in section~\ref{sec:conclusion}.

\section{Methodology}
\label{sec:methodology}

\subsection{Flow configuration}
\label{subsec:flow}

\begin{figure}
\centering
\psfrag{xx}[][][1.25]{$L_x$}
\psfrag{yy}[][][1.25]{$L_y$}
\psfrag{zz}[][][1.25]{$L_z$}
\psfrag{xt}[][][1.25]{$x_t$}
\psfrag{xsh}[][][1.25]{$x_s$}
\includegraphics[scale=1]{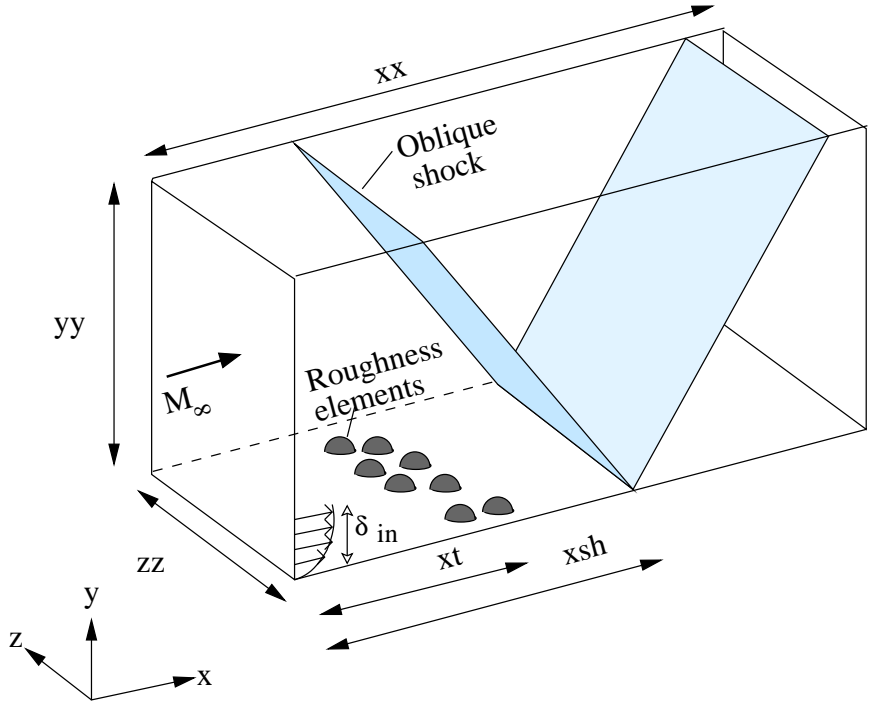}
\caption{Schematic overview of the computational setup for the DNS cases.}
\label{SCHEMATIC3D}
\end{figure}

A schematic view of the flow configuration investigated is shown in figure~\ref{SCHEMATIC3D}.
The overall size of the computational domain is $L_x \times L_y \times L_z = 120 \delta_{in} \times 35 \delta_{in} \times 10 \delta_{in}$ in the
streamwise (x), wall-normal (y) and spanwise (z) directions, respectively, $\delta_{in}$ being the thickness (based on 99 \% of the  external velocity) of the laminar
boundary layer imposed at the inflow station.
At the domain inlet, the Reynolds number based on the momentum thickness is $Re_\theta=6424$ and the freestream Mach number is $M_{\infty} = 1.7$.
A strip of roughness elements of width $4.5 \delta_{in}$ is centered around a streamwise distance $x_t=31\delta_{in}$ from the domain inlet. 
Ten hemispherical elements of height $k = 0.5 \, \delta_{95}$ are randomly distributed along the entire spanwise width. Here, $\delta_{95}$ is the
thickness (based on 95 \% of the external velocity) of the boundary layer at the trip location.  
The Reynolds number at the trip is $Re_{x_t}= 1.4 \times 10^6$, which coincides with the experimental value considered by \citet{giepman16}.
The corresponding roughness Reynolds number, i.e. the Reynolds number associated with the element height $k$ and
quantities evaluated at the roughness edge, is $Re_k = 3.1 \times 10^3$, a value sufficiently high to trigger roughness-induced transition~\citep{tani69,bernardini14}.

We run a total of five DNS, whose parameters are listed in table \ref{TAB_CASES}.
Case BL-TRIP corresponds to the simulation of a boundary layer tripped by roughness elements without any impinging shock.
The other cases include shock/boundary-layer interactions  for varying impingement location and shock strength.
We choose two values of shock strength, corresponding to flow deviations $\phi=3^o$ and $\phi=6^o$. 
For a given shock strength, we choose two points of shock impingement, $x_s/\delta_{in}=48$ and $x_s/\delta_{in}=83.8$, corresponding to (see section~\ref{sec:bl}) a
transitional- and a turbulent boundary layer, respectively.
For two of the DNS cases (BL-TRIP and SH3-TU), experimental data from~\citet{giepman16} are available for comparison.  

%TTTTTTTTTTTTTTTTTTTTTTTTTTTTTTTTTTTTTTTTTTTTTTTT
\begin{table} 
 \centering
 \begin{tabular}
 %\hline 
 {c @{\hskip 0.5cm}  c @{\hskip 0.5cm}  c @{\hskip 0.5cm} c @{\hskip 0.5cm} c @{\hskip 0.5cm} l @{\hskip 0.5cm} c @{\hskip 0.5cm}  c  }
 %\hline 
 Test case & $M_\infty$ &   $Re_\theta$     &  $\phi $ & $x_t/\delta_{in}$ & $x_s/\delta_{in}$ & Grid  & Interaction \\ 
 %\hline
 BL-TRIP &   1.7   &    6424 &    - & 31  & -  &  $4096\times 592 \times 384$ & No shock     \\
 SH3-TR  &   1.7   &    6424 &    3$^{\circ}$ & 31  & 48 &  $4096\times 592 \times 384$ & transitional \\
 SH3-TU  &   1.7   &    6424 &    3$^{\circ}$ & 31  & 83.8 &  $4096\times 592 \times 384$ & turbulent  \\
 SH6-TR  &   1.7   &    6424 &    6$^{\circ}$ & 31  & 48 &  $4096\times 592 \times 384$ & transitional \\
 SH6-TU  &   1.7   &    6424 &    6$^{\circ}$ & 31  & 83.8 &  $4096\times 592 \times 384$ & turbulent  \\
 %\hline
 \end{tabular}
\caption{Details of the parameters for the DNS cases. $M_\infty$ is the freestream Mach number, $Re_\theta$ is the 
Reynolds number based on the momentum thickness at the inlet of the computational domain, and $\phi$ is the flow deflection angle.
The distances from the leading edge to the roughness elements ($x_t$) and the shock impingement point ($x_s$) 
are also specified.} %The number in the case notation corresponds to the angle of the shock generator, $TR$ corresponds to the transitional interaction and $TU$ corresponds to the turbulent interaction.}
\label{TAB_CASES}
\end{table}
%TTTTTTTTTTTTTTTTTTTTTTTTTTTTTTTTTTTTTTTTTTTTTTT

\subsection{Numerical Method}
\label{subsec:method}

We discretize the flow domain using a Cartesian grid and solve the three-dimensional compressible Navier-Stokes equations for a perfect 
gas with Fourier heat law and Newtonian viscous terms. The fluid under consideration is air with a value of specific heat ratio 
$\gamma=1.4$, specific gas constant $R=287$ $KJ/kg^oK$ and molecular Prandtl number $\Pran=0.72$. We assume the molecular viscosity 
$\mu$ to depend on the temperature $T$ through the Sutherland's law and compute the thermal conductivity as $k_c = c_p \mu / \Pran$, 
where $c_p$ is the specific heat at constant pressure. 
We employ $4096 \times 592 \times 384$ points for discretization along the streamwise, wall-normal and spanwise 
directions, respectively, and generate a uniform grid spacing in the wall-parallel directions. In the wall-normal direction 
we cluster the grid nodes towards the wall by adopting a hyperbolic sine mapping ranging from $y = 0$ up to $y = 5 \delta_{in}$, which 
is succeeded by a uniform mesh spacing using a suitable smoothing in the connecting zone. We ensure sufficient grid refinement by evaluating 
the wall units at  $x/\delta_{in}=83.8$ (turbulent regime) for case BL-TRIP. The wall units are obtained by normalizing
 the grid spacing in terms of the viscous length scale $\delta_v$, and take the value of  $\Delta x^+ = 4.63$ and $\Delta z^+ = 4.12$ along 
the streamwise and spanwise directions, respectively. Along the wall-normal direction, the value ranges from $\Delta y^+ = 1.02$ at 
the wall to $\Delta y^+ = 21.4$ at the edge of the boundary layer.
%ATTENTION

The boundary conditions are specified as follows. At the inlet, a laminar boundary layer is imposed, whose
profile is determined from the solution of the generalized Blasius equations~\citep{white74}.
An oblique shock wave is introduced at the top of the domain by locally enforcing the inviscid
jump relations so as to mimic the effect of the shock generator.
Non-reflecting  boundary conditions are enforced at the top wall and at the outlet of
the domain to avoid spurious wave reflections. A characteristic wave decomposition is
also applied at the adiabatic no-slip wall to ensure perfect reflection of acoustic waves.

The governing equations are solved using an in-house finite-difference flow solver, widely validated for 
wall-bounded flows and SBLI in the transonic and supersonic regimes~\citep{pirozzoli10,bernardini16h}.
The solver is based on state-of-the-art numerical algorithms designed to tackle the challenging problems 
associated with high-speed turbulent flow solutions, allowing to accurately resolve a 
wide spectrum of turbulent scales and to capture steep gradients without unwanted numerical oscillations. We discretize the 
convective terms of the governing equations using a sixth-order central differencing scheme, and in the shock regions,
identified through the Ducros sensor~\citep{ducros99}, we use a fifth-order WENO scheme. To improve the numerical stability, 
the convective terms are arranged in skew-symmetric form~\citep{reiss14} and the triple split proposed by~\citet{kennedy08} is applied in a locally conservative formulation. The viscous terms are
expanded to Laplacian form and discretized using a sixth-order central differencing scheme, which guarantees physical dissipation at the
smallest scales resolved by the computational mesh.
The solution is advanced in time using a third-order, low-storage, explicit Runge-Kutta algorithm~\citep{bernardini09}. 
The presence of the roughness elements is managed by means of the immersed-boundary (IB) method, that allows to deal with
embedded geometries of arbitrary shape on a Cartesian grid. In the present study, the 
IB method is implemented following the approach proposed by~\citet{detullio_07} for compressible flows.
Additional details on the implementation can be found in~\citet{bernardini16}.

We carry out the simulations using 2048 cores on the Lenovo NeXtScale platform at the Italian computing center CINECA, using
a total budget of 2.25 Mio. CPU hours.
The flow statistics were computed over a time period of about $327 \delta_{in}/U_{\infty}$ using around 1200 flow fields.
In the present work, we normalize the streamwise distance either using the trip location i.e. $x_t^*=(x-x_t)/\delta_{95}$,
or by the inviscid shock impingement point, i.e. $x_s^*=(x-x_s)/\delta_{in}$.  

\section{Boundary-layer transition}
\label{sec:bl}

%\begin{figure}
%\begin{centering}
%\includegraphics[scale=0.30]{figs/R_LAMINAR_U_PROF_RED}
%\par\end{centering}

%\caption{The laminar Blasius profile at $x/\delta_{\text{in}}=20$ before encountering the roughness elements. The wall-normal direction is normalized by the boundary-layer thickness at the trip location, based on 0.95$U_\infty$. Also shown is the laminar boundary layer profile in the experiments of \citet{giepman16} (symbols) before encountering the distributed roughness elements.}

%\label{COMP_LAMINAR_BLASIUS}
%\end{figure}

In this section, we look at the effect of the transition device on the incoming boundary layer
in the absence of the impinging shock, and we provide a characterization of the boundary-layer
evolution along the streamwise direction.

\begin{figure}
\begin{centering}
\includegraphics[scale=0.50,angle=0]{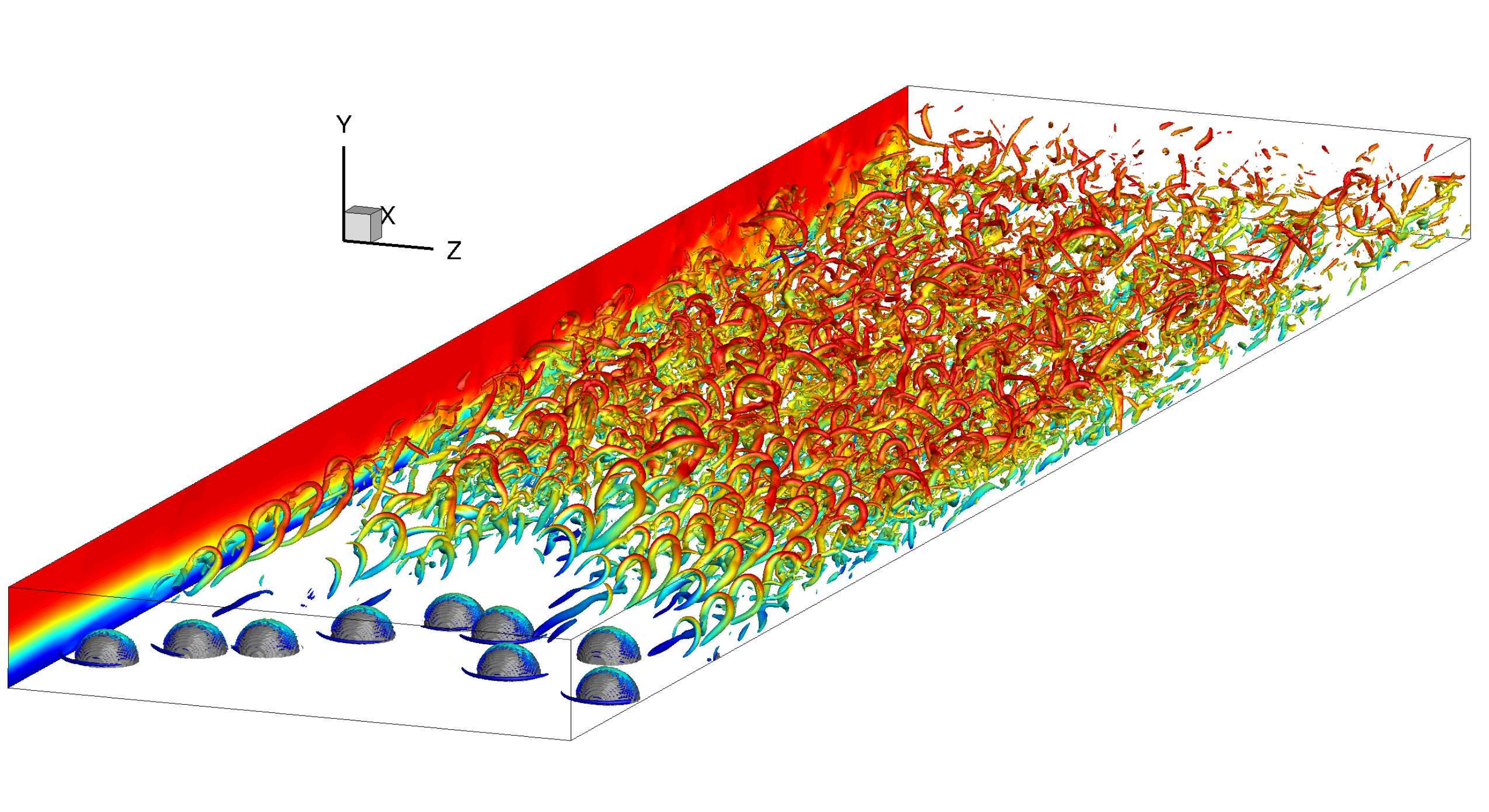}
\par\end{centering}
\caption{Three-dimensional visualization of vortical structures past the roughness elements for the case BL-TRIP
obtained as isosurfaces of the swirling strength, colored by the mean streamwise velocity using
sixty contour levels ranging from $-0.125 \leq U/U_\infty \leq 1$.}
\label{CLEAN_3D_CONTOUR1}
\end{figure}

The incoming laminar boundary layer is perturbed  after encountering the roughness elements at $x_t^* = 0$, and since the roughness height is 
above the critical value~\citep{bernardini12}, transition to turbulence occurs further downstream. Figure~\ref{CLEAN_3D_CONTOUR1}, where isosurfaces of the swirling
strength criterion are reported, shows the typical pattern to transition observed in previous studies of roughness-induced transition \citep{acarlar87,redford10},
with the formation and shedding of hairpin vortical structures past the roughness elements, which evolve in the streamwise direction leading
to the flow breakdown. This behavior is associated to the instability of the detached shear layer on the top of roughness edge and has been observed
across a wide range of Mach numbers~\citep{bernardini14}.

\begin{figure}
\begin{center}
\includegraphics[scale=0.30]{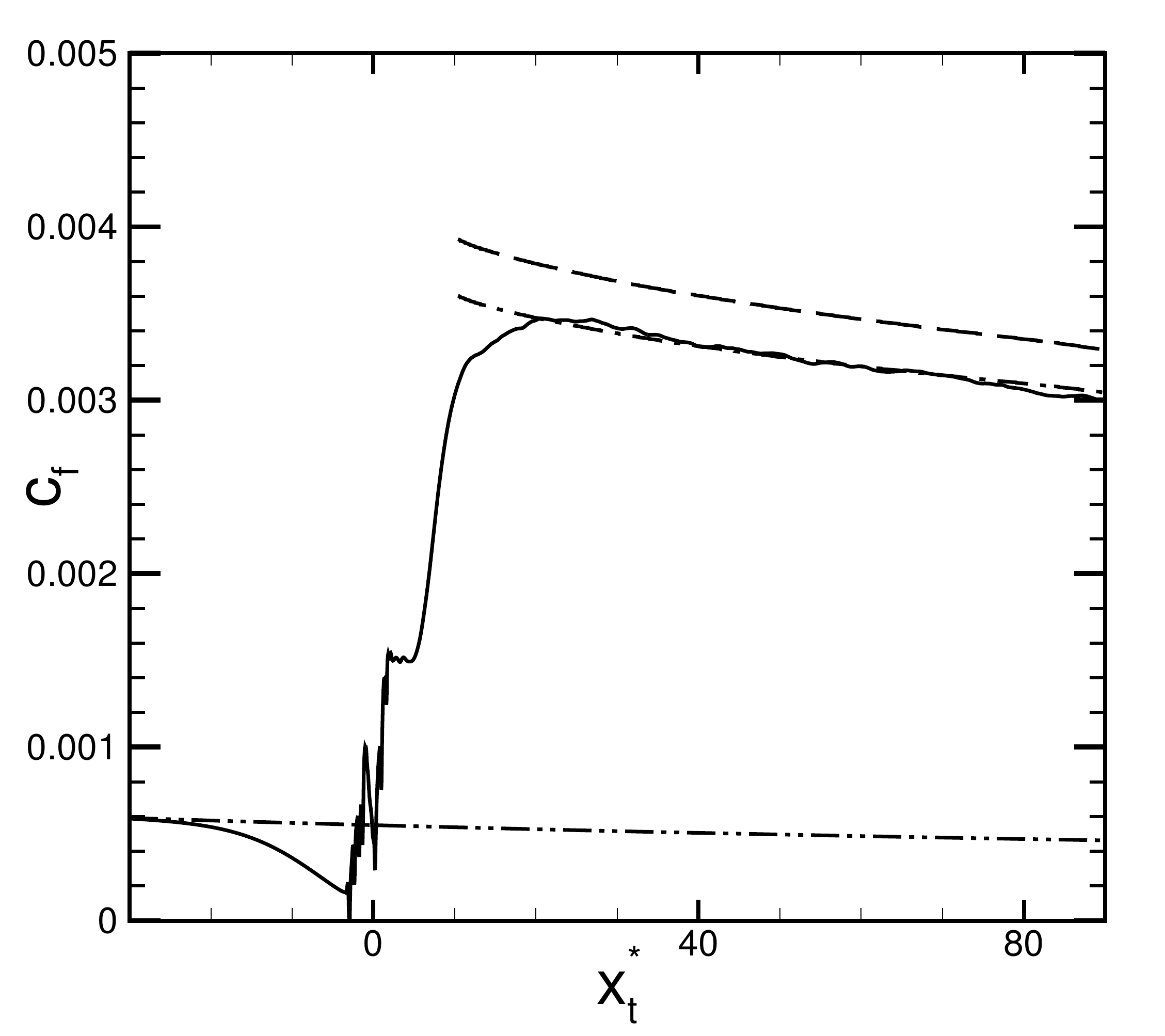}
\caption{Distribution of the mean skin friction coefficient along the streamwise direction from DNS (solid line) for the case BL-TRIP. 
Also shown are the laminar $c_f$ values (dash dot dot) and the turbulent $c_f$ values from the K\'arm\'an-Schoenherr (dash dot; 
see Eq. (\ref{KARMAN})) and Blasius (dashed;see Eq. (\ref{BLASIUS})) relations.}
\label{CLEAN_CF}
\end{center}
\end{figure}

We report the distribution of the time- and spanwise-averaged skin friction coefficient $c_f = \tau_w/q_\infty$ along
the streamwise direction in figure~\ref{CLEAN_CF},
$\tau_w$ and $q_\infty$ being the wall-shear stress and free-stream dynamic pressure, respectively.
Upstream of the roughness elements, the skin friction is lower than the corresponding laminar solution, as a consequence of the perturbation
induced by the tripping device. Beyond the roughness strip, $c_f$ increases rapidly to about seven times the corresponding laminar value and
gradually decreases further, attaining values typical of a turbulent boundary layer. For reference purpose, we also include in the figure the $c_f$  predictions
obtained from two theoretical expressions for turbulent flows, the incompressible skin-friction correlation of K\'arm\'an-Schoenherr,
\begin{equation}
c_{fi}=1/(17.08(\log_{10}Re_{\theta_i})^2 + 25.11 \log_{10}Re_{\theta_i} + 6.012),
\label{KARMAN}
\end{equation} 
and the Blasius correlation,
\begin{equation}
c_{fi}=0.026/Re_{\theta_i}^{1/4},
\label{BLASIUS}
\end{equation} 
where the subscript `i' refers to the incompressible regime.
These relations are extended to compressible flows via the van Driest \RN{2} transformation (for adiabatic wall) given by 
\begin{equation}
c_{f_{i}}=F_c c_f, \hspace{1cm} Re_{\theta_{i}}=F_\theta Re_\theta,
\end{equation} 
where 
\begin{equation}
F_c=\frac{T_w/T_\infty-1}{\arcsin^2\alpha},\hspace{1cm} F_\theta=\frac{\mu_\infty}{\mu_w}, \hspace{1cm} \alpha=\frac{T_w/T_\infty-1}{\sqrt{T_w/T_\infty(T_w/T_\infty-1)}}.
\end{equation} 
We observe a good match between the $c_f$ distribution obtained from the DNS and the K\'arm\'an-Schoenherr 
prediction for $x_t^* > 40$,
whereas the Blasius expression overpredicts the DNS result by about ten percent.

\begin{figure}
\begin{center}
\includegraphics[scale=0.40]{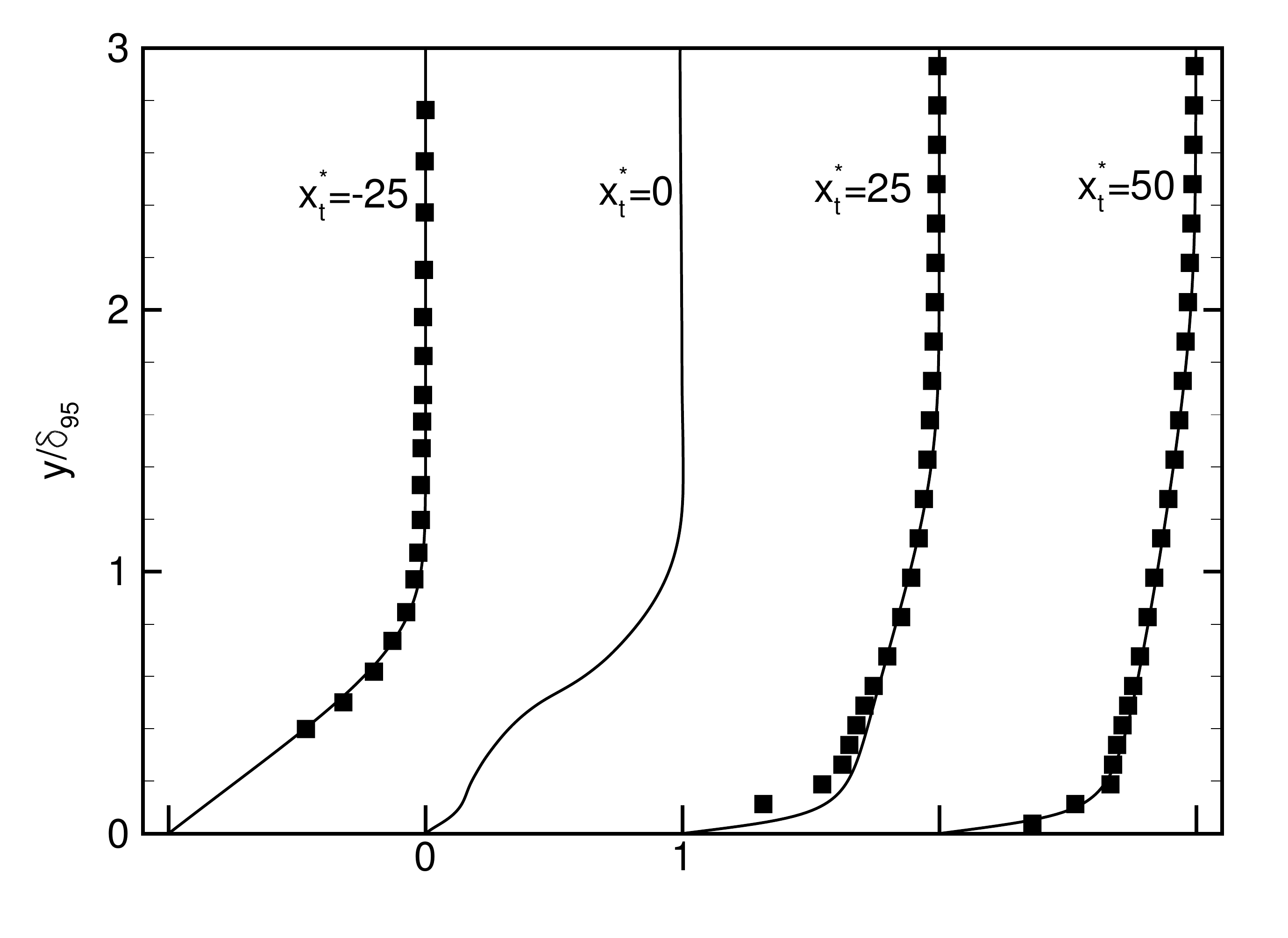}
\caption{Mean velocity profiles ($U/U_e$) along the wall-normal direction at various streamwise locations for the case BL-TRIP. Lines denote the
DNS results and symbols represent the experimental data of~\citet{giepman16}.} 
\label{R_TRIP_UPROF_VARYX}
\end{center}
\end{figure}

A comparison of the DNS data with the experiments of~\citet{giepman16}
is reported in figure~\ref{R_TRIP_UPROF_VARYX}, where mean velocity profiles at various
stations along the streamwise direction are shown.
%We report the evolution of the mean velocity profile along the streamwise direction in figure~\ref{R_TRIP_UPROF_VARYX} for various
%representative stations, and include the available experimental data of \citet{giepman16}.
The first location ($x_t^* = -25$) corresponds to the region upstream of the tripping elements, and the velocity profile is characterized by an
extended linear behavior typical of a laminar boundary layer. A strongly perturbed profile is observed at $x_t^*= 0$,
with the presence of two inflection points, which indicates the onset of the instabilities due to the flow interaction
with the tripping elements.
As a consequence of the transition process, the mean velocity profiles at subsequent stations ($x_t^*=25$ and $x_t^*=50$)
become fuller, with a steeper velocity gradient at the wall.
We obtain a very good agreement with the experimental data, which demonstrates the capability of the simulation in
accurately predicting the streamwise evolution of the flow and capturing the length-scale of the transition process.

\begin{figure}
 \centering
 a)
 \includegraphics[scale=0.28]{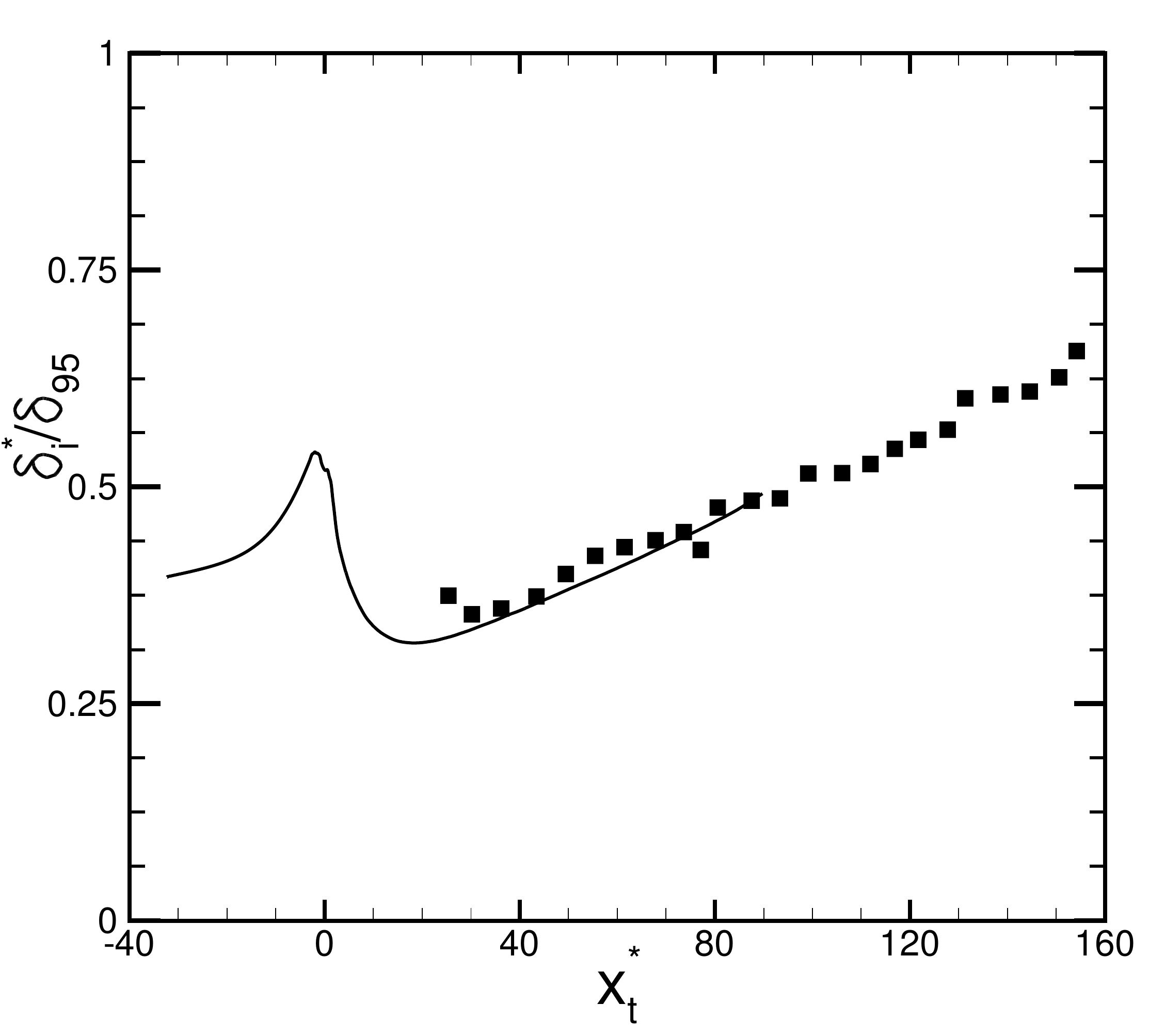} \hskip 1em
 b)
 \includegraphics[scale=0.28]{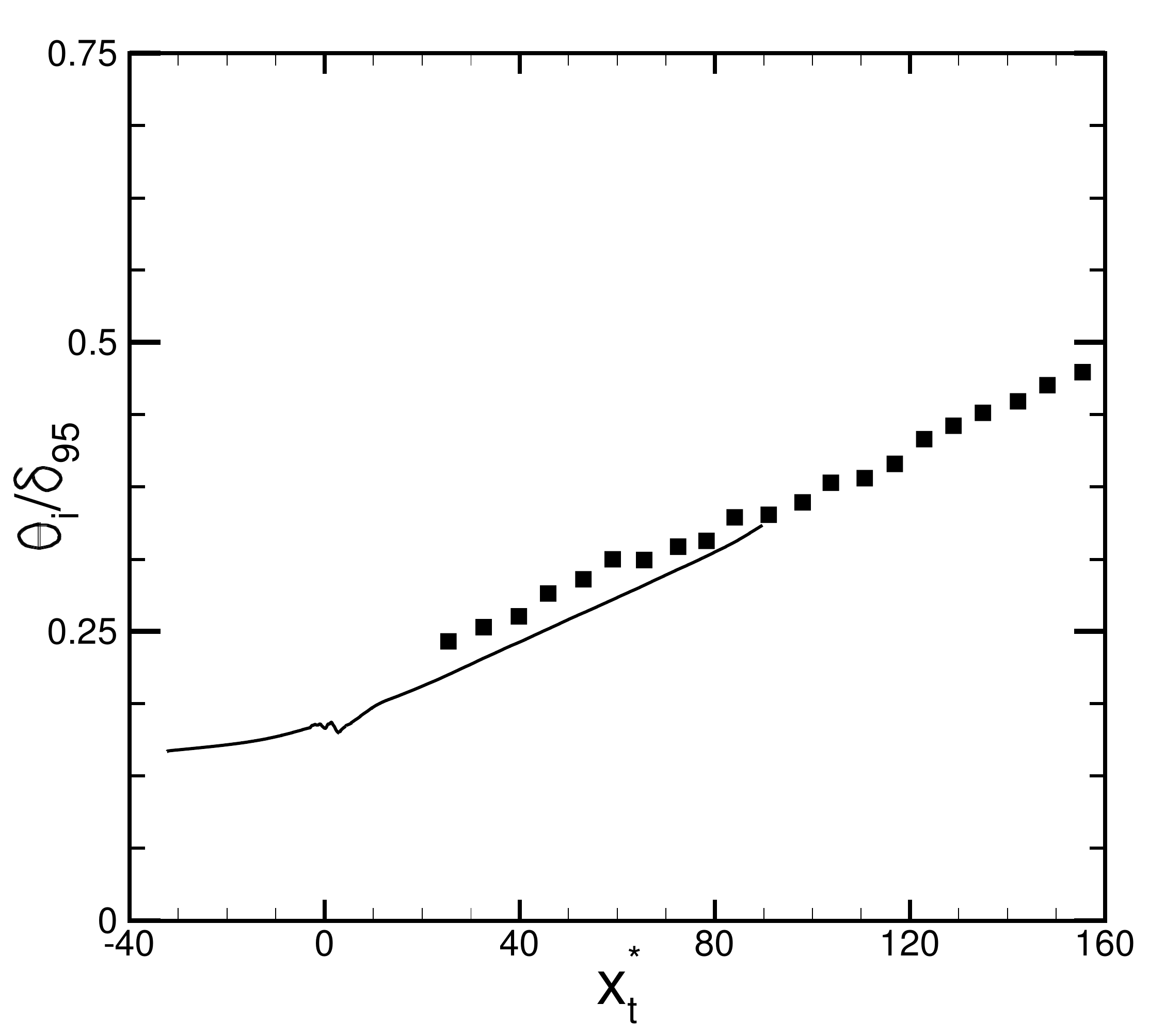} \vskip 1em
 c)
 \includegraphics[scale=0.28]{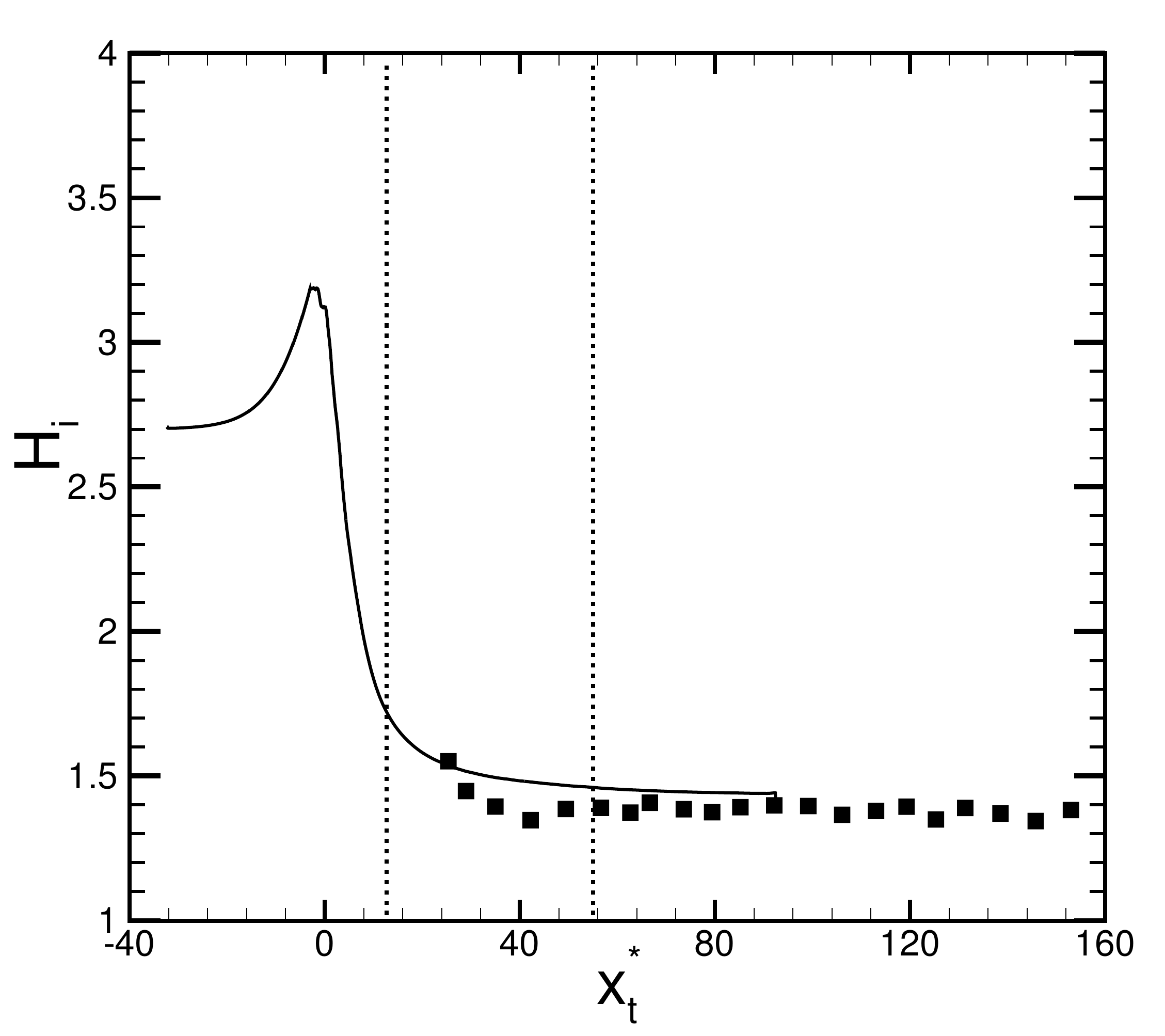} \vskip 1em
 \caption{Distribution of the incompressible (a) displacement thickness, (b) momentum thickness and (c) shape factor for the
 case BL-TRIP. Lines denote the DNS results and symbols represent the experimental data of~\citet{giepman16}.
 The two vertical dotted lines reported in panel (c) denote the location of shock impingment for the SBLI flow cases discussed in the later sections.}
 \label{CLEAN_BL}
\end{figure}

A further comparison between the DNS and the experiment is provided in figure~\ref{CLEAN_BL}, where the evolution of the boundary layer
is described in terms of incompressible displacement thickness ($\delta^*_i$), momentum thickness ($\theta_i$) and shape factor ($H_i$),
\begin{equation}
\label{thicknesses}
%\delta^*_i=\int_0^{\delta_e}\left(1-\frac{\rho}{\rho_e}\frac{U}{U_e}\right)\diff  y, \qquad
%\theta_i=\int_0^{\delta_e}\frac{\rho}{\rho_e}\frac{U}{U_e} \left(1-\frac{U}{U_e}\right)\diff  y, \qquad
%H_i = \frac{\delta_i^*}{\theta_i}.
 \delta^*_i=\int_0^{\delta_e}\left(1-\frac{U}{U_e}\right)\diff  y, \qquad
 \theta_i=\int_0^{\delta_e}\frac{U}{U_e} \left(1-\frac{U}{U_e}\right)\diff  y, \qquad
 H_i = \frac{\delta_i^*}{\theta_i}.
\end{equation} 
The agreement between the DNS results and the experimental data is very satisfactory, also considering
the challenges of performing PIV measurements in extremely thin transitional boundary layers
with non-uniform seeding distributions.
Past the tripping elements the displacement thickness initially decreases
as a consequence of the transition process that fills-up the velocity profile, it achieves
a minimum at $x^*_t \approx 15$ and then it starts to rise. On the other hand the momentum thickness is characterized
by a steady growth and its post-trip value is always larger than the value computed 
at the trip location. The distribution of the shape factor reflects the transition process undergone by the boundary layer.
It has an initial value of about 2.7, typical of a laminar boundary layer and attains a peak of about 3.2 at the roughness location.
Past the interaction, the shape factor displays a drastic drop to about 1.4, which is a value typical of a turbulent boundary layer~\citep{smits06},
achieved approximately for $x_t^* > 40$.

\section{Effect of impinging shock}
\label{sec:sh}

\begin{figure}
 \centering
 a)
 \includegraphics[trim=2 0 0 0,clip,scale=0.32,angle=-90]{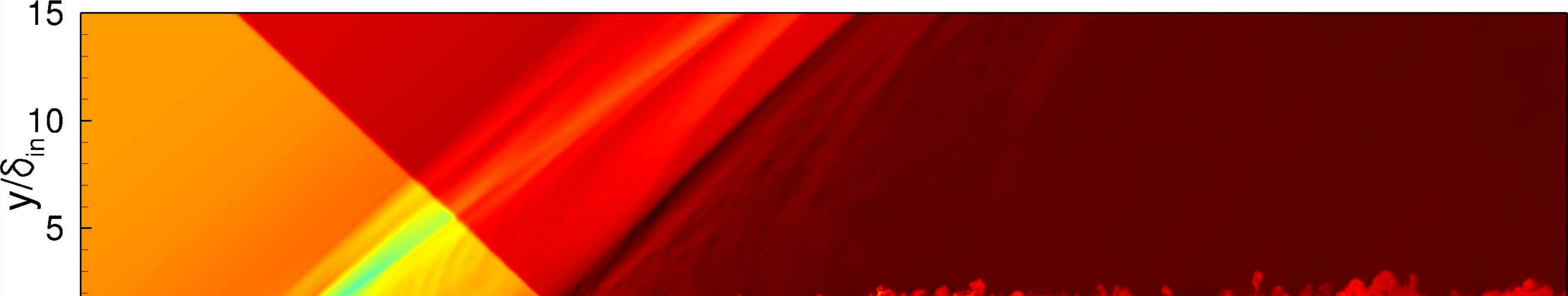} \vskip 1em
 b)
 \includegraphics[trim=0 0 0 0,clip,scale=0.32,angle=-90]{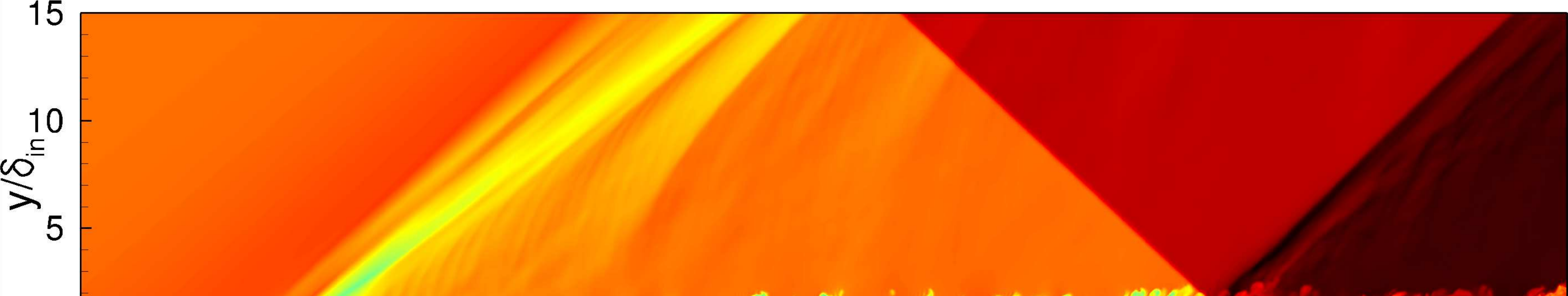} \vskip 1em
 c)
 \includegraphics[trim=2 0 0 0,clip,scale=0.32,angle=-90]{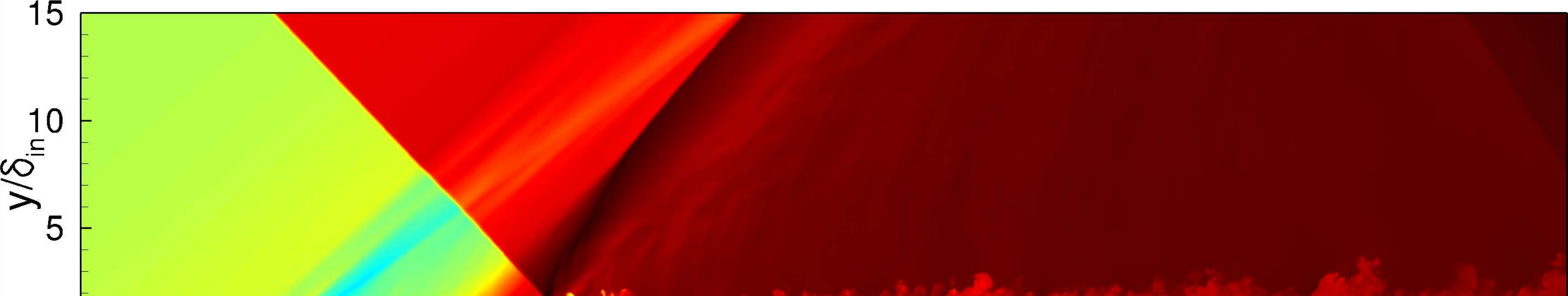} \vskip 1em
 d)
 \includegraphics[trim=2 0 0 0,clip,scale=0.32,angle=-90]{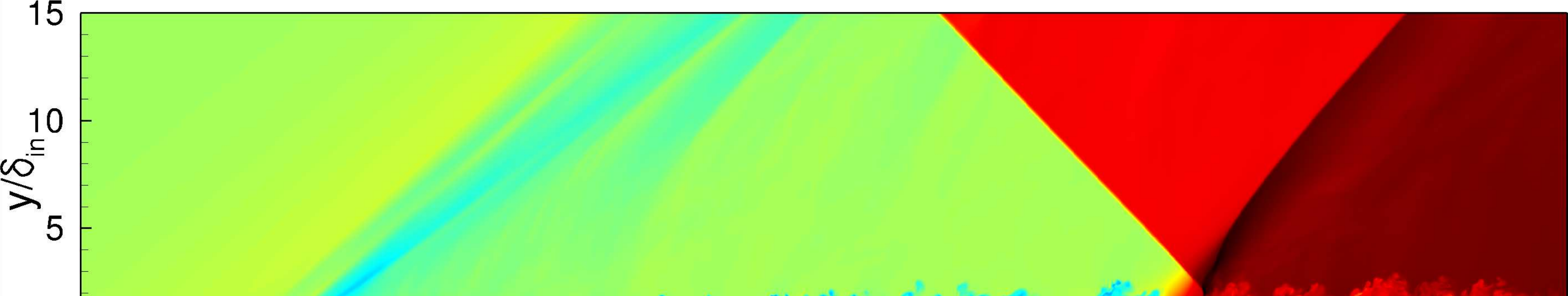} \vskip 1em
 \caption{Contours of instantaneous density for the four cases of shock impingement, 
 a) SHK3-TR, b) SH3-TU, c) SH6-TR and d) SH6-TU. 
 Fifty contour levels are shown in the range of $0.56<\rho/\rho_\infty<1.33$ for the 
 $\phi=3^o$ cases and $0.56<\rho/\rho_\infty<1.83$ for the $\phi=6^o$ cases.}
 \label{SHK_CONTOUR_2D_DENSITY}
\end{figure}

%\begin{figure}
%\begin{center}
%  \begin{subfigure}[b]{1\textwidth}
%    \includegraphics[scale=0.40,angle=0]{figs/SHK_348_3D_CONTOUR}
%    \caption{}
%    \label{SHK_348_3D_CONTOUR}
%  \end{subfigure}
%  \newline
%\hspace{1cm}
%  \begin{subfigure}[b]{1\textwidth}
%    \includegraphics[scale=0.40,angle=0]{figs/SHK_648_3D_CONTOUR}
%    \caption{}
%    \label{SHK_648_3D_CONTOUR}
%  \end{subfigure} 
%\caption{Vortical structures generated past the roughness elements identified by the swirling
%strength criterion  impinged upon by an oblique shock wave for the cases  (a) SH3-TR and
%(b) SH6-TR. The slice shows the
%contour of mean streamwise velocity, which is also  used color the vortices.}
% \label{SHK_3D_CONTOUR}
%\end{center}
%\end{figure}

%\begin{figure}
%\begin{centering}
%
%
%\includegraphics[scale=0.50,angle=0]{figs/SHK_348_3D_CONTOUR}
%\par\end{centering}
%\caption{Vortical structures generated past the roughness elements identified by the swirling 
%strength criterion  impinged upon by an oblique shock wave for case 2. The slice shows the 
%contour of mean streamwise velocity, which is also  used color the vortices.}
%\label{SHK_348_3D_CONTOUR}
%\end{figure}

In this section we investigate the interaction of an impinging shock with the spatially 
evolving boundary layer discussed in the previous section, with the main aim of characterizing 
the effect of the incoming boundary-layer state (transitional or turbulent) on the properties of 
the interaction.
To that purpose, we analyze the results of the four DNS listed in table~\ref{TAB_CASES},
performed for two values of shock strength ($\phi=3^o$ and $\phi=6^o$) and two shock impingement
locations $x_s/\delta_{in} = 48 $ and $83.8$ ($x_t^* = 17.7$ and $55$), corresponding to
transitional and turbulent SBLI, respectively. This classification is supported by the
results discussed in the previous section (see in particular figure \ref{CLEAN_BL}c).

To provide a qualitative overview of the flow organization, we report
in figure~\ref{SHK_CONTOUR_2D_DENSITY} contours of the instantaneous density
in a longitudinal $x-y$ plane. The density field highlights very well the
wave system originated as a consequence of the interaction process,
mainly consisting of the impinging and the reflected shock, as well as a series of waves
(compression-expansion-compression) radiating in the freestream,
arising from the perturbation of the boundary layer induced by the tripping device.
We point out that the other feature typically observed in a strong SBLI involving a separation bubble,
a fan of expansion waves associated with the reattachment of the boundary layer,
is not observed in figure~\ref{SHK_CONTOUR_2D_DENSITY} and
both the transitional and turbulent interaction cases share the same qualitative behavior.
This reveals that the present interactions do not involve a massive
separation of the flow, contrary to the case of an untripped laminar
oblique shock-wave reflection at $\phi = 3^\circ$, considered under the same flow conditions (Mach- and Reynolds numbers)
in the experimental work by~\citet{giepman15}, who highlighted the presence of a large separation bubble.
A major role is played by the strength of the impinging shock, which determines the
extent of the interaction zone, that significantly increases with the deflection angle $\phi$.
%We present a detailed discussion into the characteristic interaction length in the subsequent text.  

%We next investigate the effect of an impinging shock on the transitional and turbulent boundary layer.
%The four cases here analyzed are those reported in table~\ref{TAB_CASES}, performed for two values
%of the shock strenght ( $\phi=3^o$ and $\phi=6^o$) and two impingment shock locations
%$x_s/\delta_{in} = 48 $ and $83.8$ corresponding to transitional and turbulent SBLI,
%respectively.
%
%ATTENTION !! Add details for instantaneous density figure. 

%To provide a qualitative overview of the flow organization, we report
%in figure~\ref{SHK_3D_CONTOUR} a three-dimensional instantaneous visualization
%of the vortical structures in the interaction zone for cases SH3-TR and SH6-TR,
%coloured according to the local value of the streamwise velocity.
%The figures highlight the highly three-dimensional character of the interaction, with the predominance
%of hairpin-shaped vortex tubes in the interaction region. The adverse pressure gradient imposed by the shock
%induces a mild lift off of the boundary layer, which is particularly evident for the
%stronger interaction.

\begin{figure}
 \centering
 a)
 \includegraphics[scale=0.28]{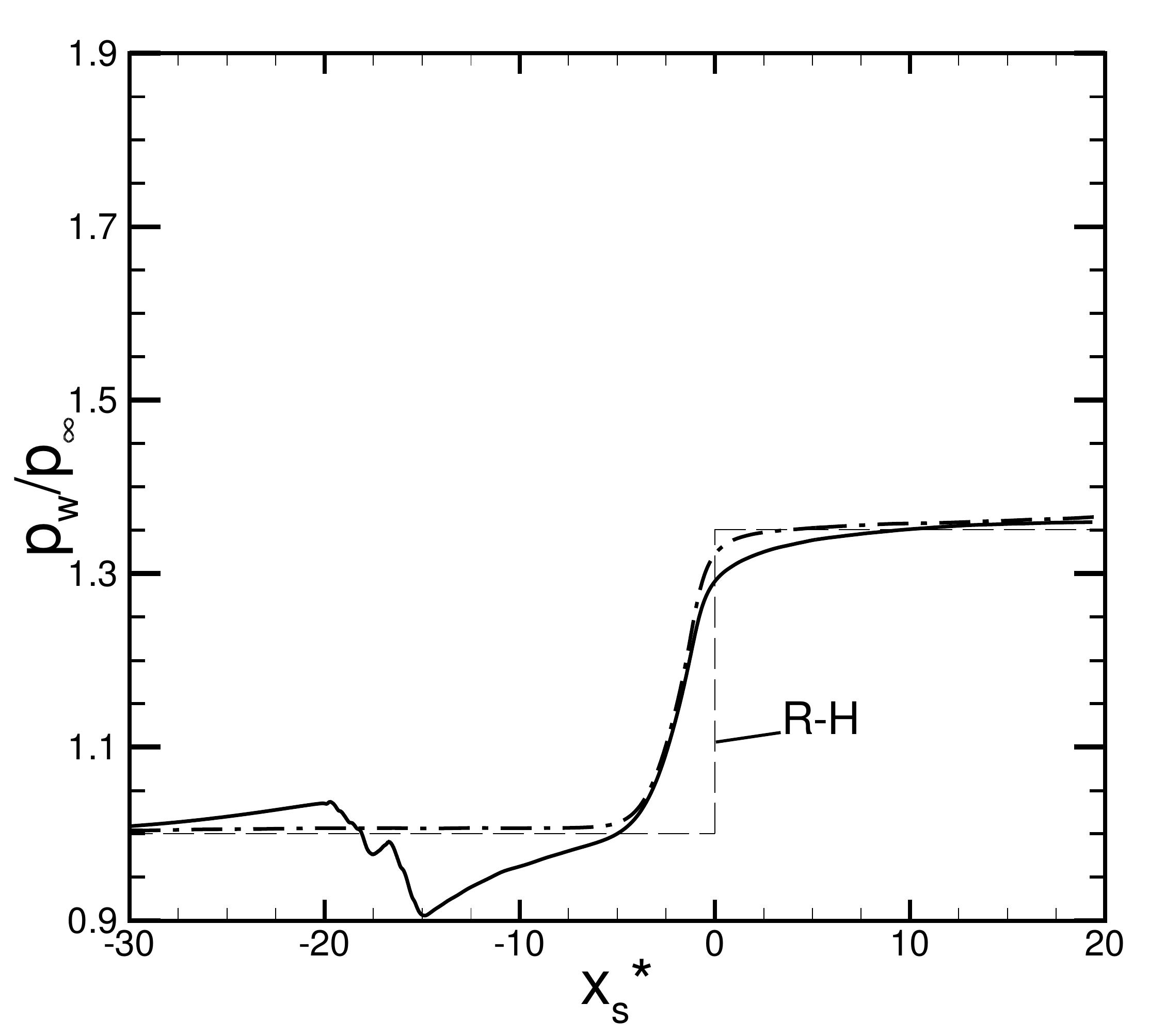} \hskip 1em
 b)
 \includegraphics[scale=0.28]{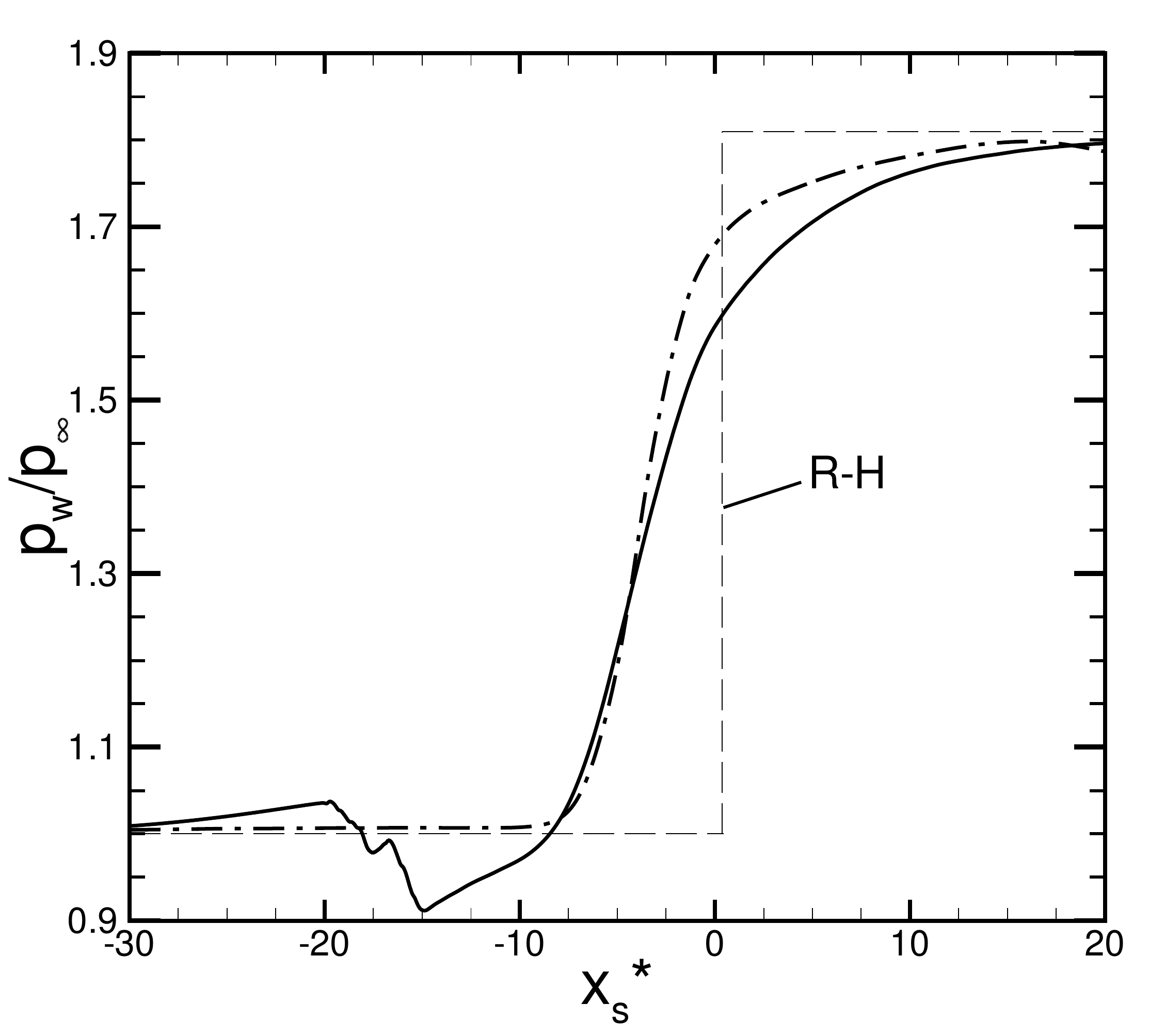} \vskip 1em
 \caption{Streamwise distribution of mean wall pressure for (a) $\phi=3^o$ and (b) $\phi=6^o$.
 Solid lines refer to transitional interactions and dash-dot lines denote turbulent interactions.
 $R-H$ denotes the inviscid distribution resulting from the Rankine-Hugoniot 
 jump conditions (dashed).}
 \label{SHK_PWALL}
\end{figure}

The distribution of the mean wall pressure is shown in figure~\ref{SHK_PWALL},
together with the inviscid pressure jumps predicted by the Rankine-Hugoniot relations.
In all cases, the wall pressure exhibits a sharp rise upstream of
the nominal shock impingement point, followed by a slower increase further downstream,
which is more gradual in the case of the transitional SBLI.
On the other hand, we observe that the beginning of the interaction
is rather independent of the nature of the incoming boundary layer
(transitional or turbulent), and the extent of the upstream influence
region is determined by the shock strength.

\begin{figure}
 \centering
 \includegraphics[scale=0.30]{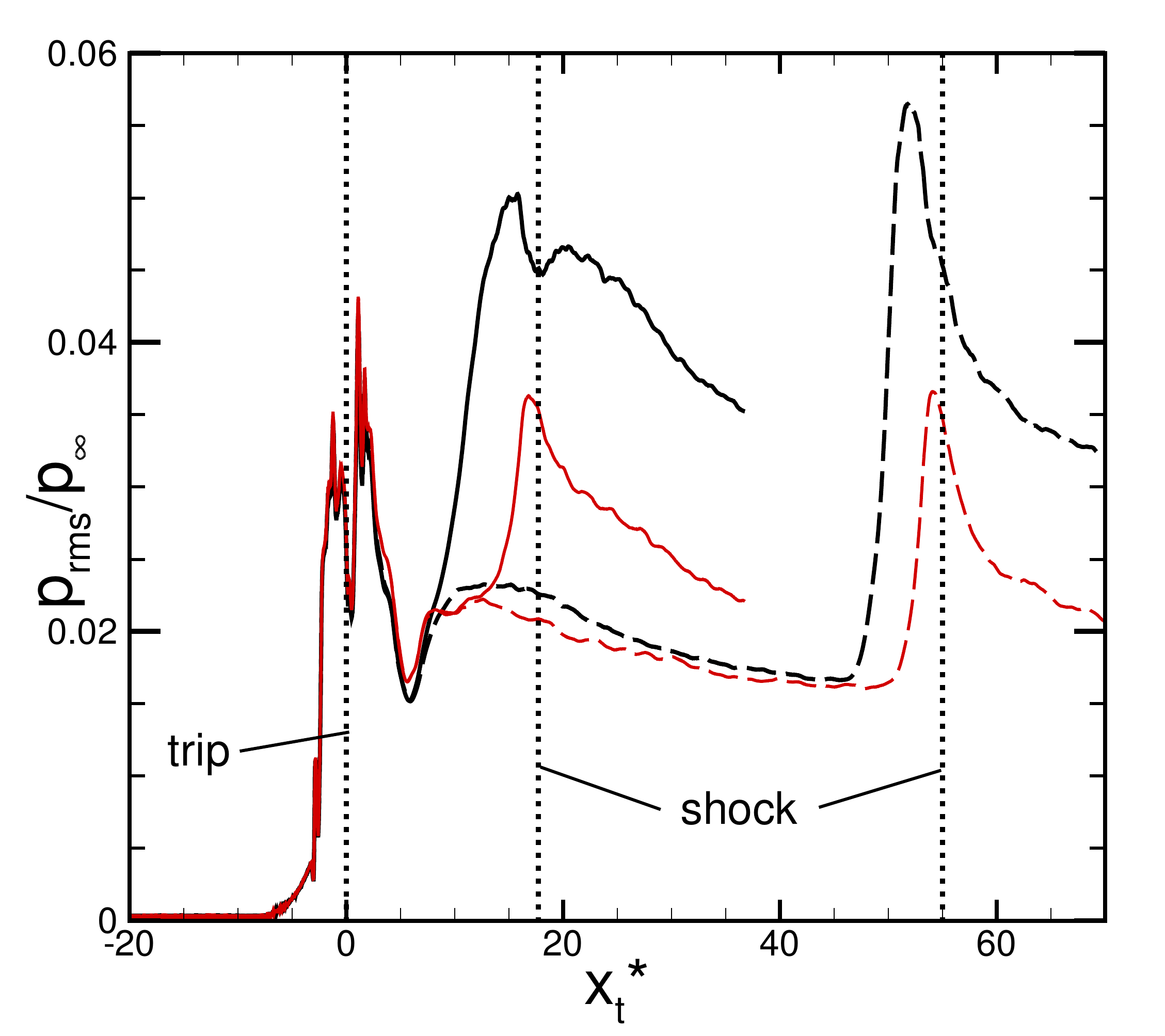} \vskip 1em
 \caption{The streamwise variation of the rms wall pressure for the $\phi=3^o$ (red) and the $\phi=6^o$
 (black) cases with the transitional (solid) and turbulent (dashed) interactions.}
 \label{SHK_PRMS}
\end{figure}

Figure \ref{SHK_PRMS} shows the root-mean-square of wall pressure 
fluctuations ($p_{w_{rms}}$) for all shock-impingement cases.
The $p_{w_{rms}}$ rise across the interaction region is sligthly higher for the turbulent interactions
as compared to the corresponding transitional cases, for both the incident shock angles.
The post-shock decay for the two types of interaction also varies, with the turbulent 
cases displaying a steeper decline past the interaction region compared to the
corresponding transitional case. 
The transitional interactions have a broader post-shock decay region, with the 
$\phi = 6^{\circ}$ case also displaying a local minimum at the nominal shock impingement point. 
%Notice that the roughness elements introduce significant level of wall pressure 
%fluctuations, even larger than what is observed in the shock-impingement region for the 
%$\phi=3^o$ cases. 
%We present the wall root-mean-square values of the pressure fluctuation ($p_{w-rms}$) for all the cases 
%of shock-impingement in figure \ref{SHK_PRMS}. 
%We observe a large value of $p_{w-rms}$ in the vicinity of the tripping device ($x_t^*=0$), 
%even larger than what is observed in the shock-impingement region for the $\phi=3^o$ case. 
%The peak value of $p_{w-rms}$ for the $\phi=6^o$ case is higher in the turbulent interaction 
%as compared to the transitional interaction by about $10/%$.
%At this shock strength, the transitional interaction also reveals a drop and rise in the 
%$p_{w-rms}$ value at the nominal shock impingement point, whereas the turbulent interaction 
%reveals a  rapid post-shock evolution associated with acoustic decay .   
%

\begin{figure}
 \centering
 a)
 \includegraphics[scale=0.28]{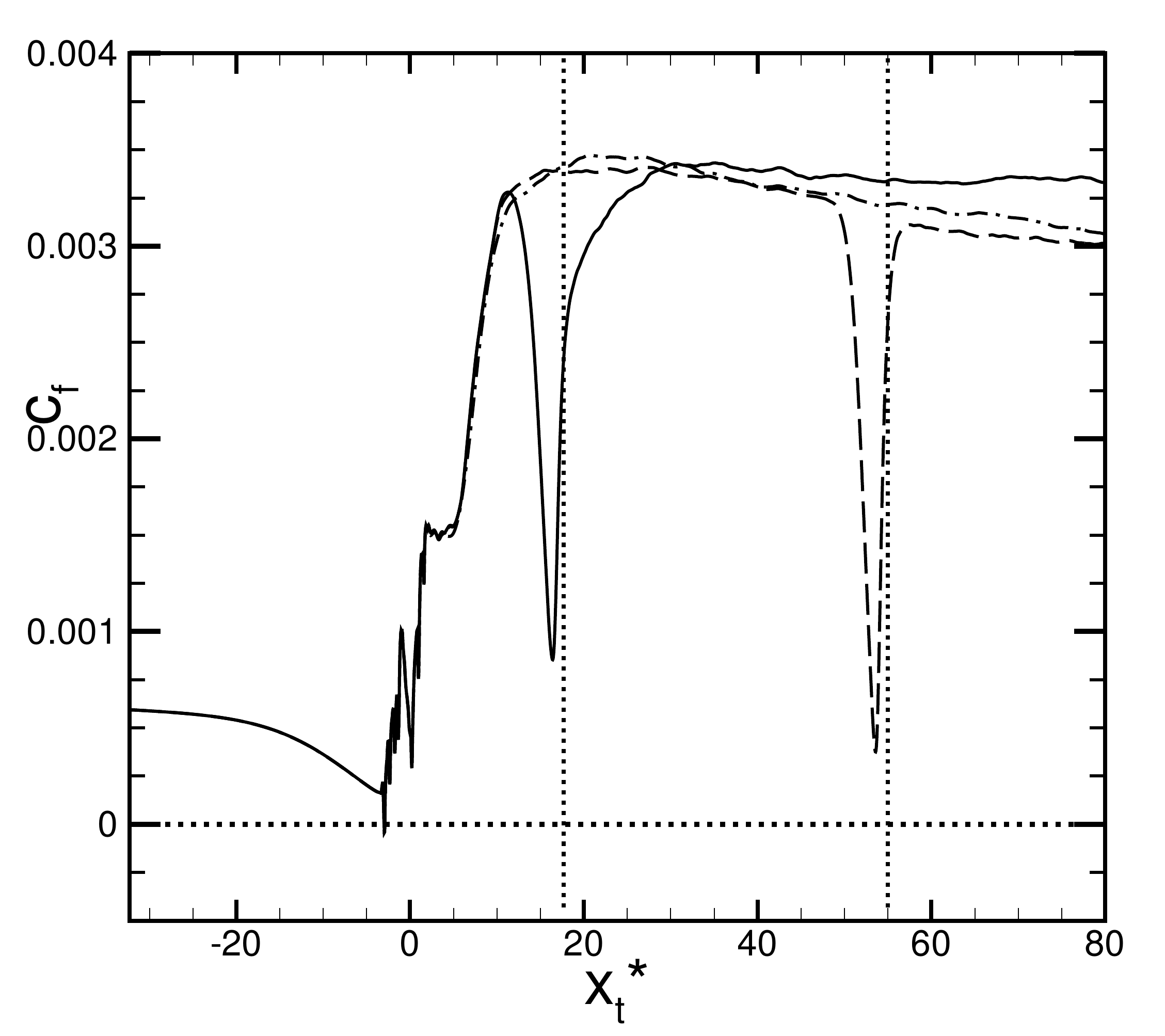} \hskip 1em
 b)
 \includegraphics[scale=0.28]{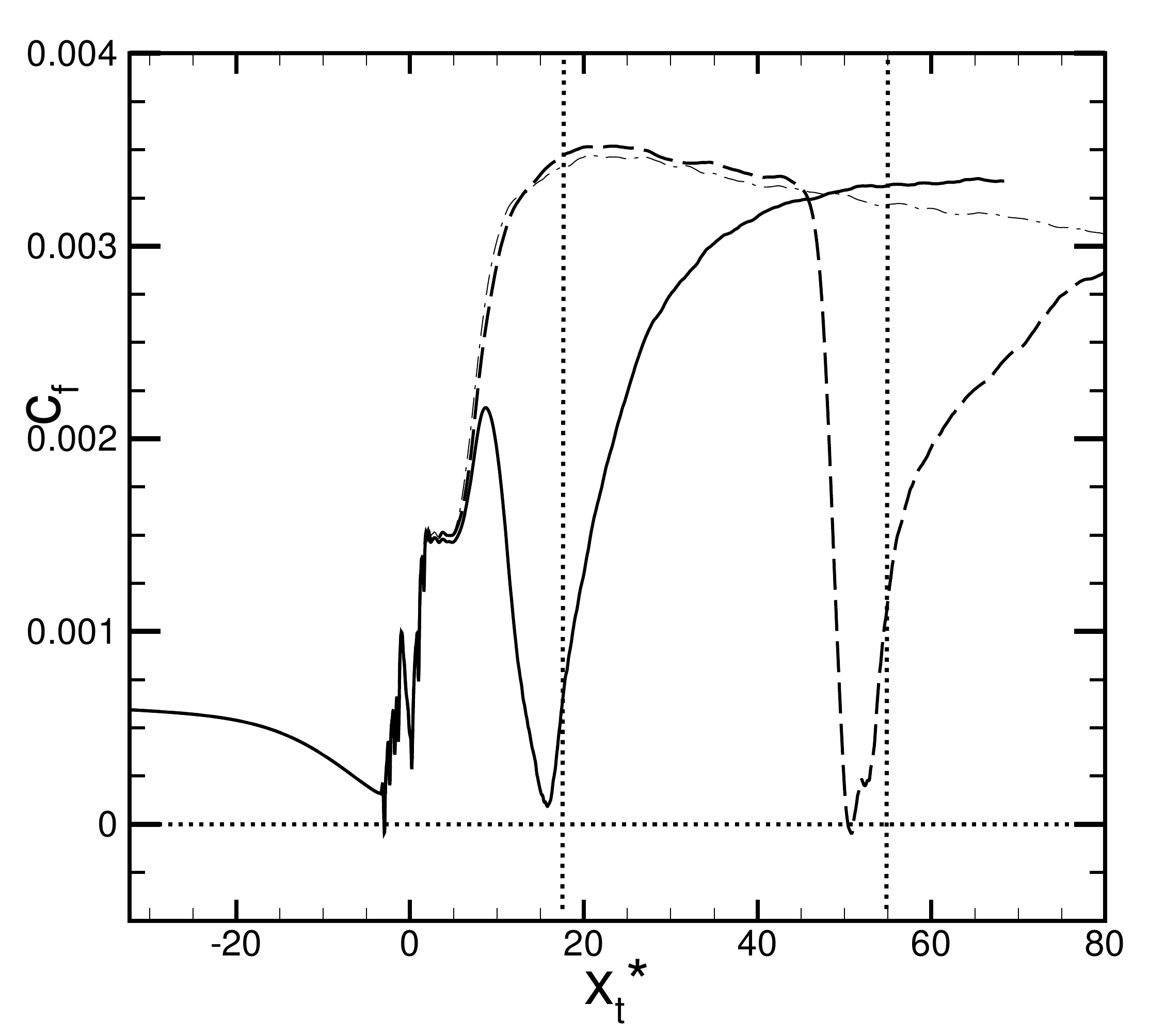} \vskip 1em
 \caption{Distribution of skin friction coefficient along the streamwise direction for the shock
 impingement cases at (a) $\phi=3^o$ and (b) $\phi=6^o$.
 Solid lines refer to the transitional interactions and dashed lines denote the turbulent interactions.
 The skin friction coefficient for the case without shock impingement is also shown (dash-dot).
 The vertical dotted lines the nominal shock impingment location.}
 \label{SHK_CF}
\end{figure}

The effectiveness of boundary-layer tripping in suppressing a shock-induced separation can 
be identified by the distribution of the skin friction coefficient $c_f$,
displayed in Fig~\ref{SHK_CF} for all the shock impingement cases.
In the interaction region the wall shear stress is characterized by a remarkable drop,
associated with the lift off of the boundary layer, followed by a gradual recovery.
The $c_f$ curves show that in all the cases, even those at higher shock strength, mean separation
is not observed, and tripping the boundary layer eliminates the large separation bubble found
in a laminar interaction. This quantitatively confirms our expectations drawn from the
inspection of figure~\ref{SHK_CONTOUR_2D_DENSITY}. The skin friction levels remain well above the zero line for
the cases SH3-TR and SH3-TU, whereas they are almost tangent for the cases SH6-TR and SH6-TU,
which can be classified as cases with incipient separation.
Quite surprisingly, the minimum value of the skin friction is lower in the turbulent interaction
than in the transitional cases, although the width of the region where the $c_f$ value drops
due to the shock impingement is wider in the transitional interactions.

\begin{figure}
 \centering
 a)
 \includegraphics[scale=0.28]{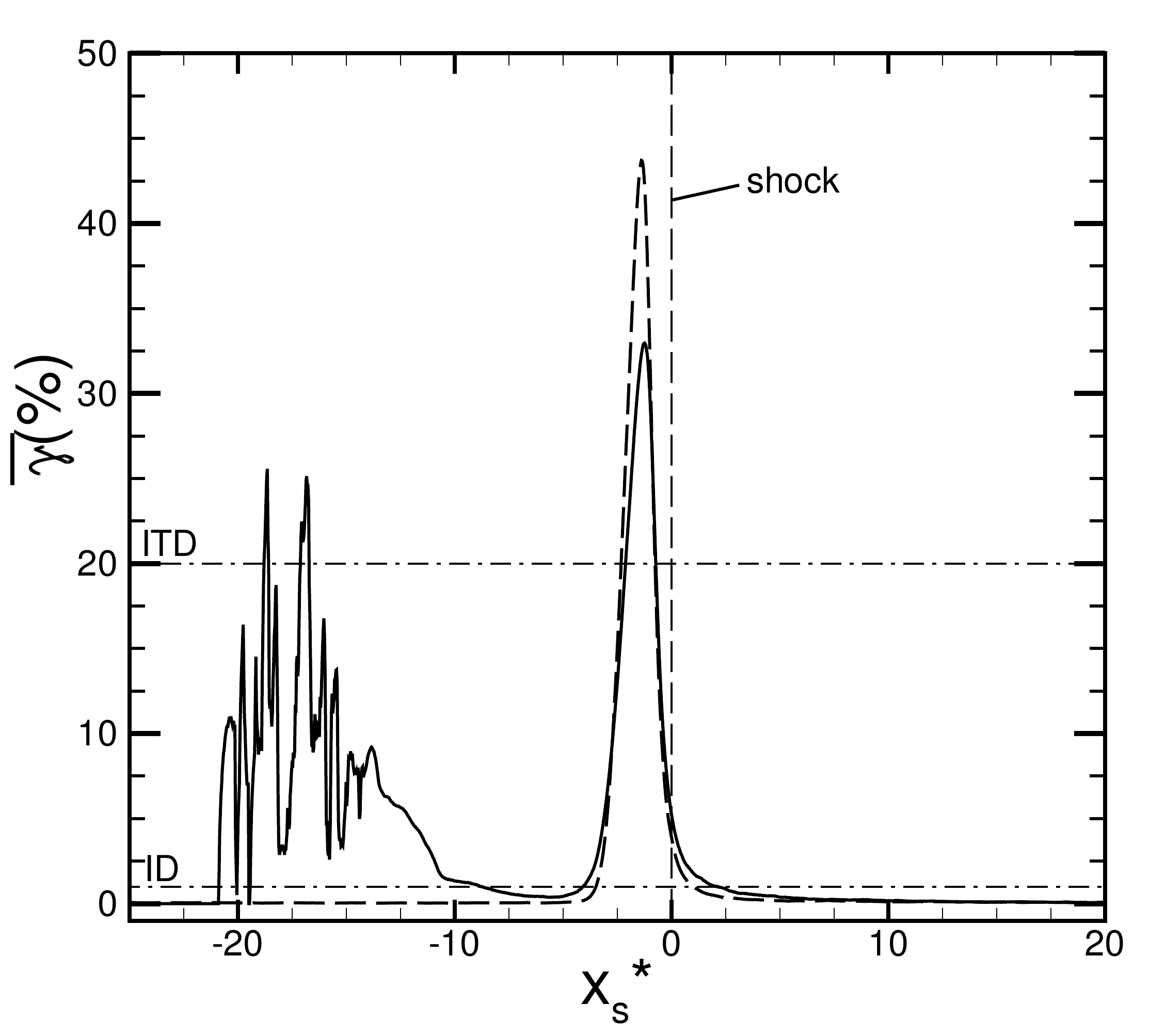} \hskip 1em
 b)
 \includegraphics[scale=0.28]{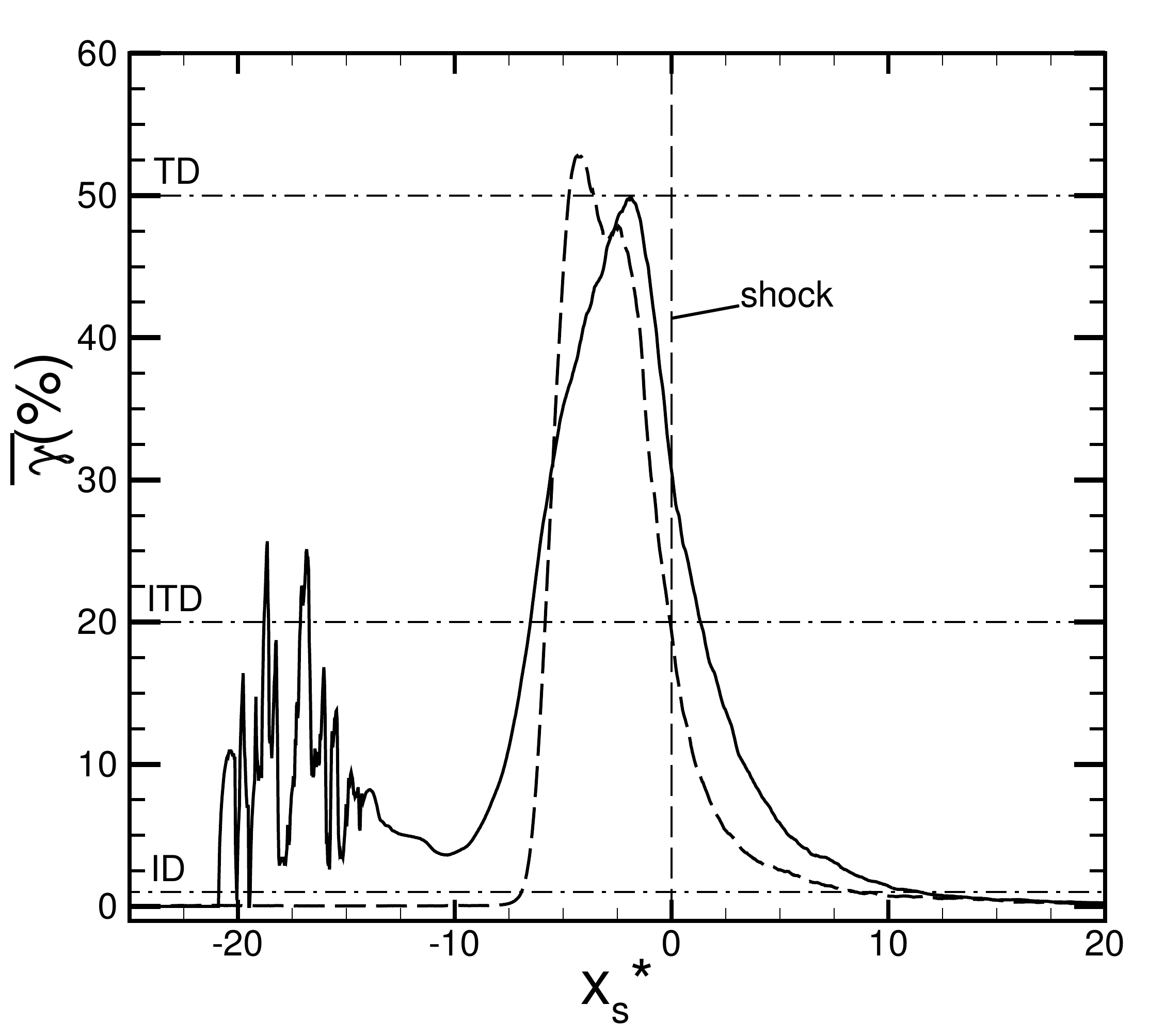} \vskip 1em
 \caption{Distribution of statistical probability of wall points along the streamwise direction with 
 $\partial u/\partial y < 0$ for (a) $\phi=3^o$ and (b) $\phi=6^o$, for the transitional interaction
 cases (solid lines) and the turbulent interaction cases (dashed lines).
 The horizontal lines denote the incipient detachment (ID), intermittent transitory detachment 
 (ITD) and the transitory detachment (TD) levels.}
 \label{SHK_STAT_PROB}
\end{figure}

The absence of a mean separation clearly does not preclude the possibility of instantaneous
zones with locally reversed flow. We characterize the regions of instantaneous separation by
plotting the statistical  probability ($\overline{\gamma}$) of the the wall points with negative
$\partial u/\partial y$, where $u$ is the instantaneous streamwise velocity.
\citet{simpson89} classified the boundary-layer detachment based on how frequently the backflow 
occurs. A statistical probability of backflow of 1\% of the total sampling is denoted as incipient 
detachment (ID), while that amounting to 20 \% is classified as intermittent transitory detachment 
(ITD). A 50 \% probability of backflow is termed as transitory detachment (TD), which clearly 
indicates a separation in the mean.

Figure \ref{SHK_STAT_PROB}a shows the statistical probability of instantaneous 
separation for the interactions at $\phi=3^o$.  
In the region of shock impingement ($x_s^*=0$), both the flow cases (SH3-TR and SH3-TU) exceed the ITD level, and the 
peak probability of separation for the turbulent interaction is slightly higher than that of the transitional 
interaction. The same plot for the flow cases at higher angle of incidence $\phi=6^o$ is shown in figure 
\ref{SHK_STAT_PROB}b. As expected, we observe higher levels of instantaneous separation 
in comparison to the lower incidence cases, with the turbulent interaction exceeding the 
transitory detachment level.
In the transitional interaction case, although the peak percentage of separation is not as high, 
the width of the separation region exceeds that of the turbulent interaction zone by about 
36 \% at the ITD level.

\begin{figure}
 \centering
 a)
 \includegraphics[scale=0.4]{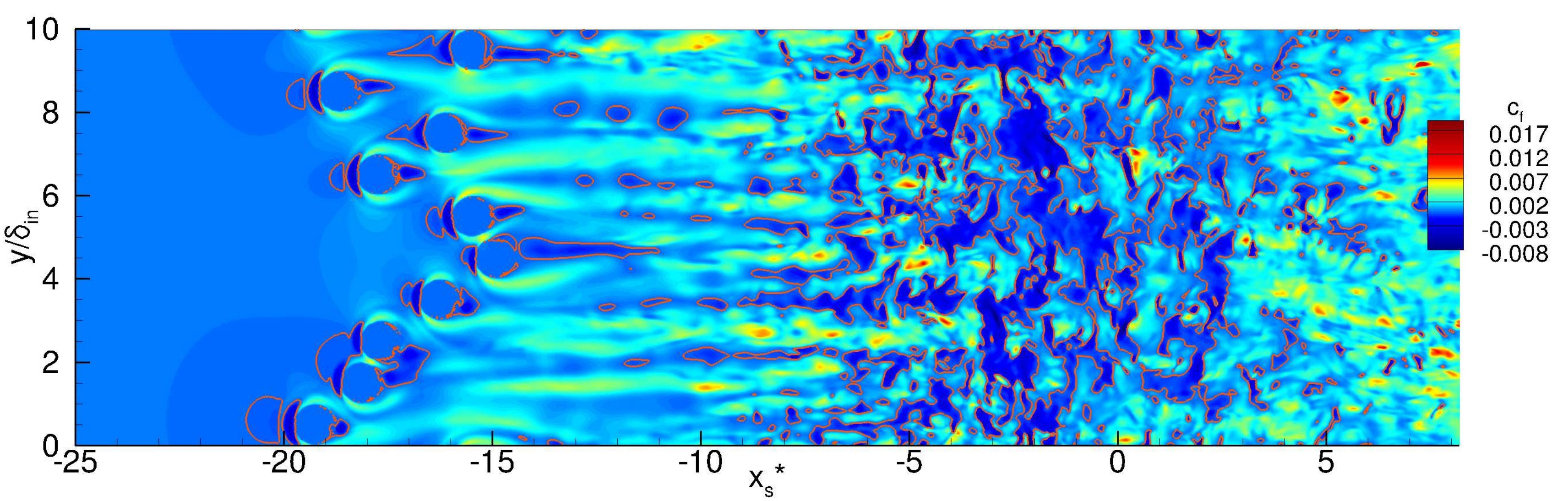} \vskip 1em
 b)
 \includegraphics[scale=0.4]{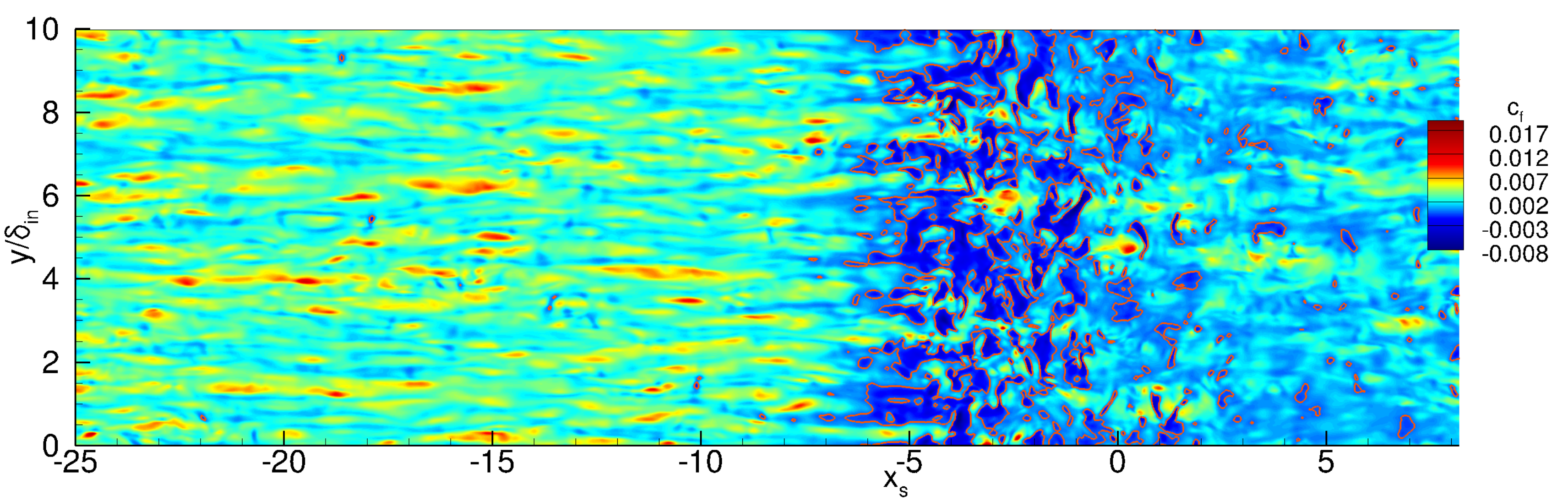} \vskip 1em
 \caption{Contours of instantaneous skin friction for the flow cases (a) SH6-TR and (b)
 SH6-TU. The iso-line $c_f=0$ is denoted in orange.}
 \label{SHK_WALL_CF}
\end{figure}

The instantaneous flow separation can be clearly visualized by plotting the wall skin friction contour,
as carried out for the $\phi=6^o$ shock impingement cases in figure \ref{SHK_WALL_CF}.
The isolines  correspond to zero skin friction level indicating areas of reverse flow.
In agreement with the statistical probability of instantaneous separation discussed previously, the
region of separation is slightly wider in the transitional interaction case as compared to the 
turbulent one.
\begin{figure}
 \centering
 a)
 \includegraphics[scale=0.28]{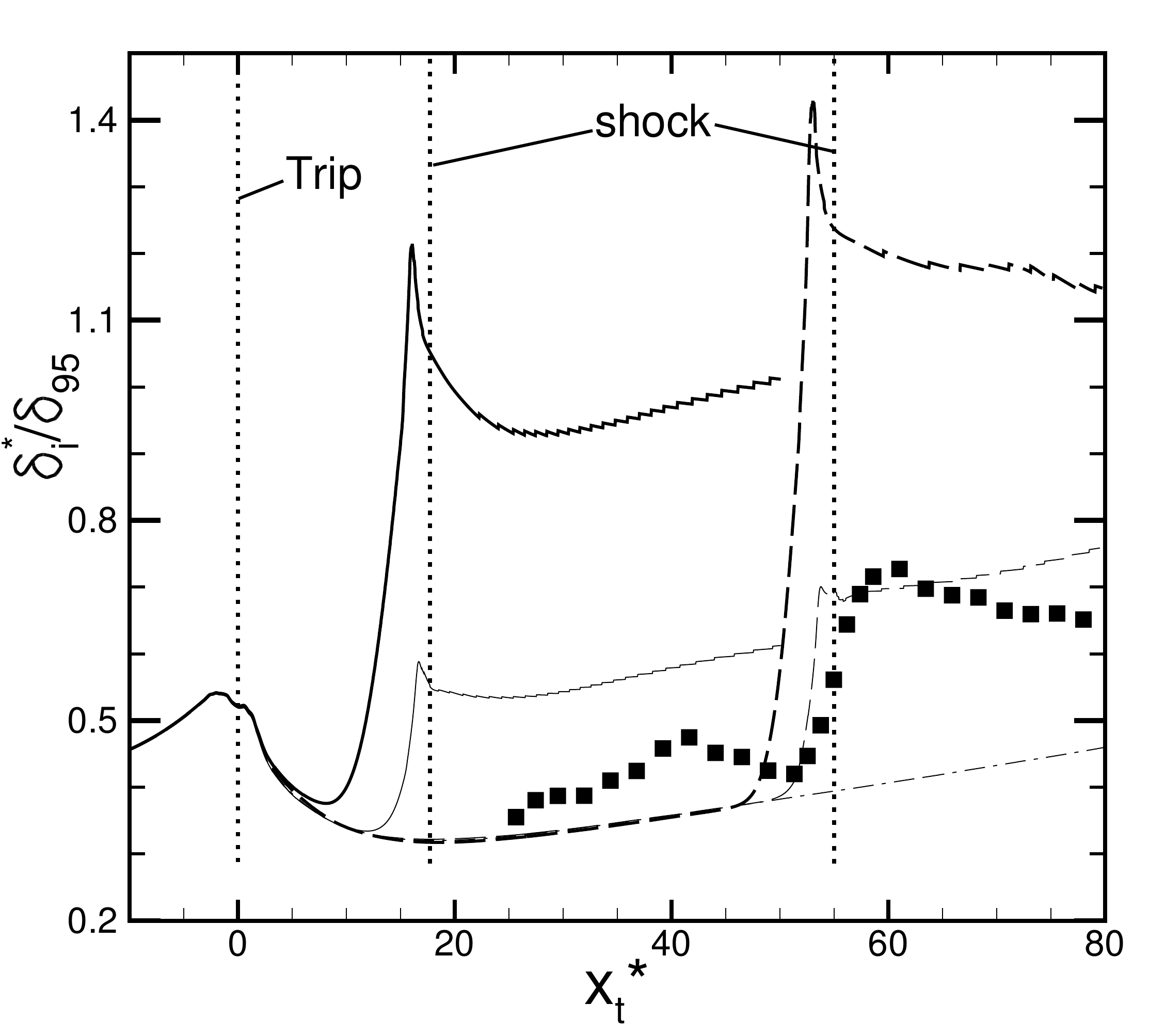} \hskip 1em
 b)
 \includegraphics[scale=0.28]{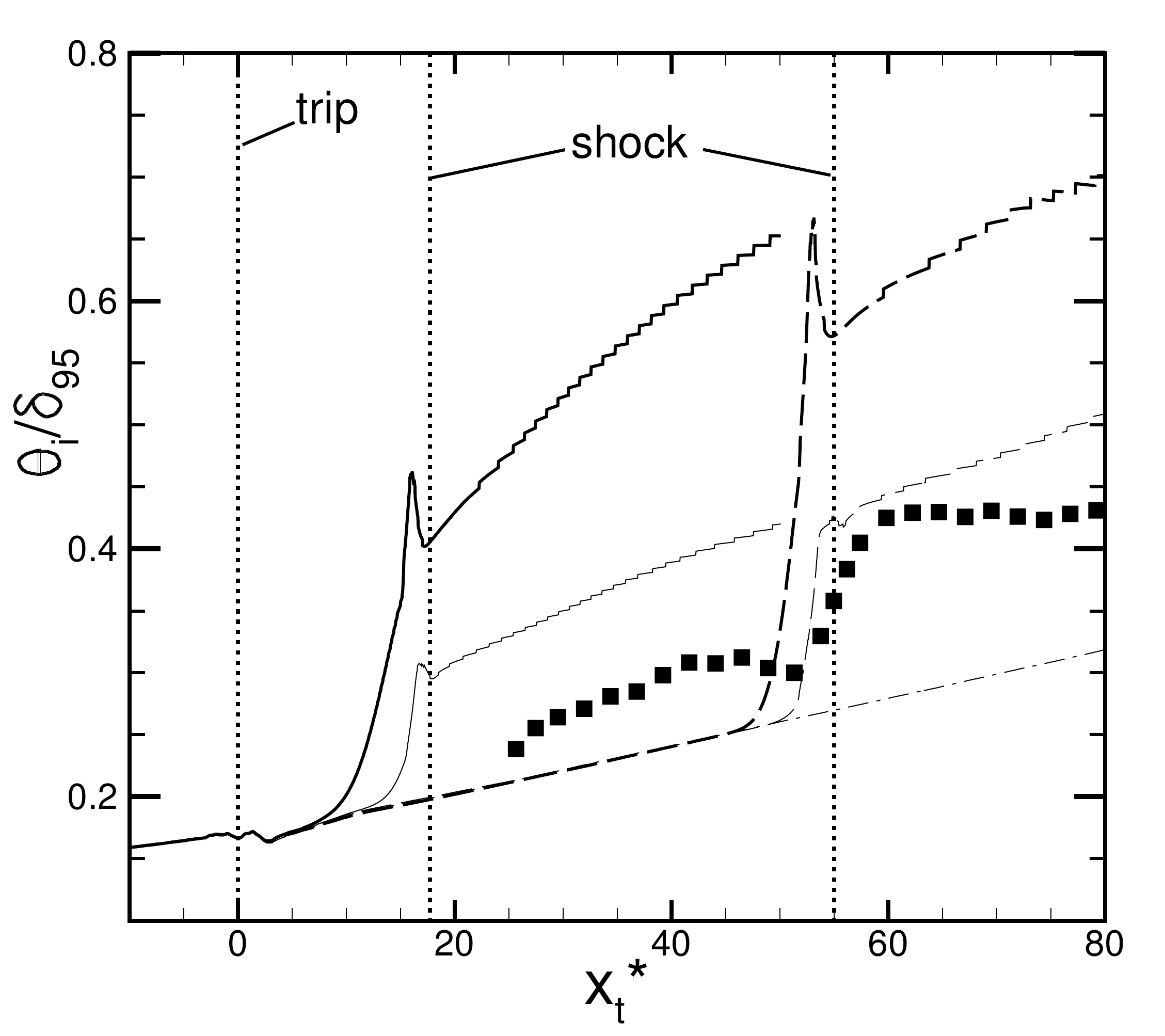} \vskip 1em
 c)
 \includegraphics[scale=0.28]{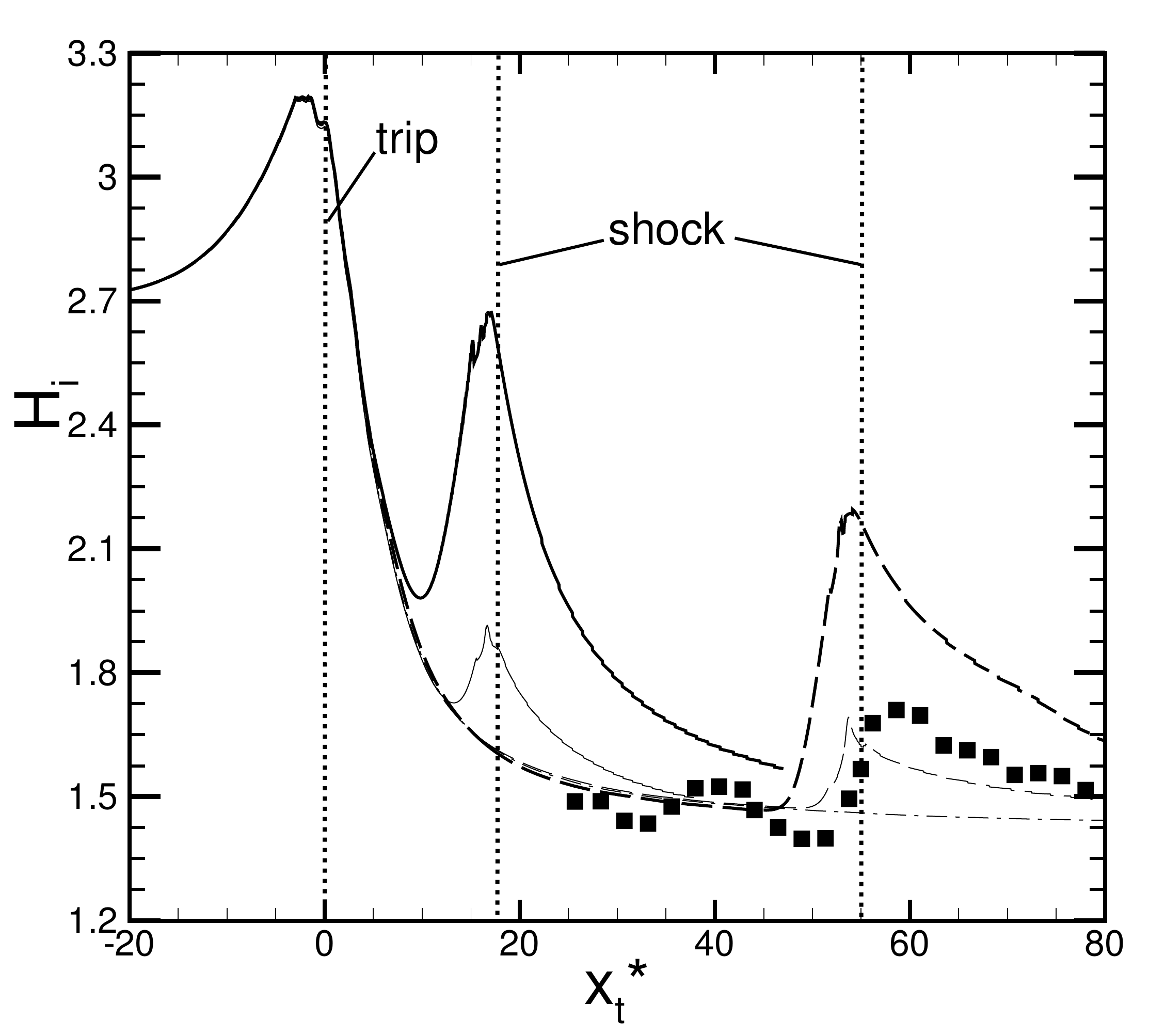} \vskip 1em
 \caption{Distribution of the incompressible (a) displacement thickness,
 (b) momentum thickness and (c) shape factor for all the shock impingement cases.
 Lines denote DNS results. Symbols refer to experiments of~\citet{giepman16}.}
 \label{SHK_6_BLT}
\end{figure}

The development of the boundary layer across the interaction is described in figure~\ref{SHK_6_BLT},
where we provide distributions of the incompressible displacement thickness,
momentum thickness and shape factor. As a reference, we also plot in the figure
the DNS results for the transitional boundary layer without any shock and
the experimental data corresponding to the flow case SH3-TU.
For all the interactions, the boundary-layer thicknesses show a rapid increase in the
region of shock impingement, which is remarkably higher for the flow cases at $\phi=6^o$ angle of incidence.
While the shock strength plays an important role, the boundary-layer growth does not seem greatly affected
by the state of the incoming boundary layer, the jump of the momentum and displacement thickness being
approximately the same for the corresponding interactions.
For the SH3-TU case, the agreement with the experimental data is fair, especially for the prediction of the location and
amplitude of the jump of $\delta_i$ and $\theta_i$. The discrepancies observed upstream of the interaction,
where the experiments exhibit a spurious peak, can be attributed to the aero-optical distortion
in the measurement and  does not reflect the actual flow field \citep{giepman16}.
The distributions of $H_i$ match quite well and highlight the boundary-layer distortion in the interaction
region, with the velocity profile becoming emptier due to the effect of the adverse pressure gradient,
leading to higher values of the shape factor.

%\subsubsection{Interaction length scales}
%TTTTTTTTTTTTTTTTTTTTTTTTTTTTTTTTTTTTTTTTTTTTTTTT
\begin{table}
 \begin{center}
 \begin{tabular}
%\hline 
{c @{\hskip 0.2cm}  c @{\hskip 0.2cm}  c @{\hskip 0.2cm}  c @{\hskip 0.2cm}  l @{\hskip 0.3cm}
c @{\hskip 0.2cm}  c @{\hskip 0.2cm}  c @{\hskip 0.3cm} c @{\hskip 0.3cm} c @{\hskip 0.3cm}
c @{\hskip 0.3cm} c @{\hskip 0.3cm}  l}
%\hline 
Test case & $M_\infty$    & $\phi$  & $\beta$ & Interaction & $\delta^*/\delta_{in}$ & $L/\delta_{in}$
& $L^*$ & $Re_\theta$ & $\overline{k}$  & $P_3/P_1$ & $S_e^*$  & Inference\\
\hline
SH3-TR & 1.7   &   3 & 38.9 & transitional & 0.32 & 3.2 & 0.56 & 8150 & 3 & 1.35 & 0.51 & attached\\
SH3-TU & 1.7   &   3 & 38.9 & turbulent & 0.36 & 3.5 & 0.55 & 12450 & 2.5 & 1.35 & 0.44  & attached \\
SH6-TR & 1.7   &   6 & 42.1 & transitional & 0.41 & 6.1 & 1.74 & 7240 & 3 & 1.81  & 1.11  & insipient Sep. \\
SH6-TU & 1.7   &   6 & 42.1 & turbulent & 0.35 & 6.1 & 2.04 & 12140 & 2.5 & 1.81  & 1.02  & insipient Sep.\\
%\hline
\end{tabular}
\end{center}
\caption{Scaling analysis of transitional and turbulent SBLI. $L^*$ is the dimensionless value of the interaction length scale,
 $\beta$ is the shock angle and $S_e^*$ the separation criterion proposed by~\citet{souverein13}.}
\label{TAB_CASES1}
\end{table}
%TTTTTTTTTTTTTTTTTTTTTTTTTTTTTTTTTTTTTTTTTTTTTTT

We now look at some of the characteristics of the interaction in the
transitional and turbulent regime at varying shock angles.
An important defining parameter for any SBLI is the
characteristic interaction length $L$, which is defined as the distance between
the wall-extrapolated point of the reflected shock and the nominal impinging shock location.
%This characteristic length is also useful in expressing the non-dimensional
%frequency of the separation shock $f$ in terms of Strouhal number \citep{dussage06} as
%\begin{equation}
%S_L=\frac{fL}{U_1}.
%\label{strouhal}
%\end{equation}

Based on a mass-balance analysis, \citet{souverein13} proposed a non-dimensional form of
the interaction length scale applicable for turbulent SBLI with adiabatic wall boundary
conditions (both, oblique shock impingement and compression corner),
\begin{equation}
L^*=\frac{L}{\delta_{in}^*}G_3,
\end{equation}
where $G_3$ is $\sin(\beta)\sin(\phi)/\sin(\beta-\phi)$, $\beta$ is the shock angle.
This new dimensionless interaction length also
classifies the interactions as attached, incipiently separated or fully separated, based on its value.
While a value of $L^*\downarrow 1$ corresponds to an attached flow, the cases with incipient
separation have $1<L^*<2$, and the separated interactions have $L^*$ values larger than two.

We provide the values of $L^*$ for our cases with shock impingement in table \ref{TAB_CASES1}.
For the two cases at $\phi=3^o$
(SH3-TR and SH3-TU), the values of $L^*$ are less than unity, and correspond to
attached flow, as shown by the skin-friction results presented in figure \ref{SHK_CF}a.
For the cases at $\phi=6^o$, $L^*=1.74$ associated with the transitional interaction
(SH6-TR) correctly classifies it as an incipient separation case.
The borderline value $L^*=2.04$ associated with the flow case SH6-TU, 
while strictly classifies it as a separated interaction, is indicative of
of an incipient separation as displayed by figure \ref{SHK_CF}b. 

\citet{souverein13} also proposed an additional parameter to characterize the SBLI in
terms of the ratio in the pressures before ($P_1$) and after ($P_3$) the shock system.
This non-dimensional parameter is given by
\begin{equation}
S_e^*=\frac{2\overline{k}}{\gamma}\frac{\frac{P_3}{P_1}-1}{M_\infty^2},
\end{equation}
and can be written as a function of free-stream Mach number $M_\infty$, flow deflection
angle $\phi$ and specific heat ratio $\gamma$. The constant $\overline{k}$ as observed from
the experimental data can either take a value of about 3 for $Re_\theta\leq 1\times 10^4$ or
a value of about 2.5 for $Re_\theta>1\times 10^4$, where $Re_\theta$ is the Reynolds number
based on the momentum thickness upstream of the interaction. The values of  $Re_\theta$,
and the associated values of $\overline{k}$ for each of the cases are listed in table \ref{TAB_CASES1}.

\begin{figure}
\begin{center}
\includegraphics[scale=0.40]{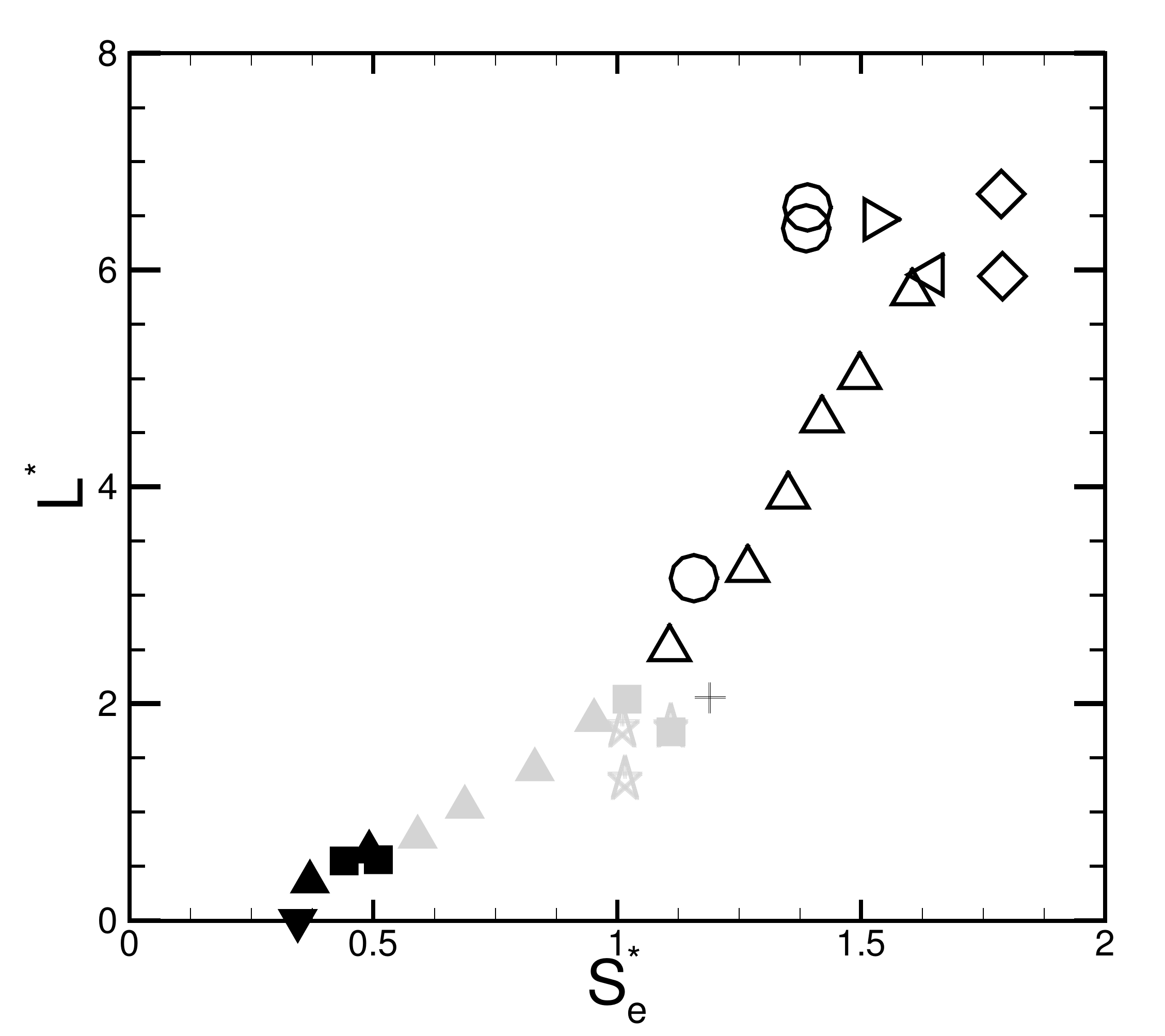}
\caption{Scaling of the interaction length. Separation criterion $S_e^*$ plotted against the interaction length $L^*$ for
the DNS flow cases (squares) compared with the available experimental data.
%Colors denote attached (black), incipiently separated (gray) and separated (white) interactions.  
Refer to~\citet{souverein13} for the interpretation of symbols.}
%Also shown is the curve fit to the experimental data (line).}
\label{SHK_SOUVEREIN_CURVE}
\end{center}
\end{figure}
According to the classification of \citet{souverein13}, a value of $S_e^*<1$ corresponds to
attached interactions and  $S_e^*>1$ represents interactions with boundary-layer separation.
The values of $S_e^*$ for each of our cases is given in table \ref{TAB_CASES1}.
%This classification holds good for both of the $\phi=3^o$ cases, where $S_e^*<1$.
%The interactions associated with $\phi=6^o$ incur a borderline value of $S_e^*>1$, and can again be 
%classified as incipient separation cases.
%Note that the criteria suggested by \citet{souverein13} was formulated for
%shock/turbulent boundary-layer interactions and some more investigation may be warranted before
%applying it to interactions involving transitional boundary layers.
\citet{souverein13} suggested that the interaction length $L^*$ plotted in combination
with the separation criterion $S_e^*$ would collapse the data along a 
single trend line irrespective of Reynolds number, Mach number and interaction type
(oblique shock impingement and compression corner).
This scaling is tested for our flow cases in figure~\ref{SHK_SOUVEREIN_CURVE},
where DNS data are reported with available experiments, and colors are used to identify
attached (black), incipiently separated (grey) or separated (white) interactions. 
We observe that, for all flow cases, our DNS data follow the experimental trend and fall within
the acceptable range of scatter, suggesting that the scaling analysis proposed by~\citet{souverein13} for turbulent SBLI,
could be equally applied to describe transitional interactions.

\section{Conclusions}
\label{sec:conclusion}

We have performed a series of direct numerical simulations to investigate the effect of  
an oblique shock wave impinging on transitional and turbulent boundary layers
at $M_{\infty} = 1.7$, with the main aim of evaluating the effectiveness of a
transitional boundary layer to suppress shock-induced separation.
%The main scope of this work was to highlight the effectiveness of a transitional boundary layer 
%in suppressing shock induced separation region, while still retaining an overall lower 
%skin friction drag as compared to a turbulent SBLI.
%This numerical study is probably the first of its kind to investigate the transitional 
%boundary-layer interaction for supersonic  inlet  Mach number  $M_\infty=1.7$ and Reynolds 
%number based on the inlet momentum thickness $Re_\theta$=6424.

The incoming laminar boundary layer was tripped by a strip of distributed roughness elements, which 
enabled a rapid transition to turbulence. A single DNS carried out without the presence of any
impinging shock wave helped to characterize the boundary-layer transition region and to validate
the numerical approach by means of a favorable comparison with available experimental data
in terms of mean velocity profiles, boundary layer thicknesses and shape factor. 
%A very good match between the available experimental data and the simulation in terms of 
%streamwise velocity profiles and the boundary layer thicknesses affirms the effectiveness 
%of the DNS in predicting the boundary-layer transition.

%Previous experiments carried out at the same Mach and Reynolds number revealed the presence 
%of a large separation bubble for a laminar (untripped) shock reflection at $\phi=3^o$.
Four DNS cases were considered in the present study based on varying shock impingement locations 
along the streamwise distance (corresponding to transitional or turbulent interactions),
and also based on varying shock strength (flow deflection angles $\phi=3^o$ and $\phi=6^o$).
We observed a clear suppression of shock-induced mean separation in both the transitional 
and turbulent interaction cases, inferred by the distribution of the skin friction coefficient.
This observation applies to both the weak as well as the strong interaction cases considered 
in our study. A higher peak in the probability of the instantaneous separation was observed for the 
turbulent interactions, although the transitional cases exhibited wider regions of 
instantaneous separation.
%A good agreement with available experimental data (in terms of boundary-layer thicknesses) 
%for one of the shock-impingement cases is encouraging. 

%Applying the criteria of  \citet{castillo01}, we identify the regions of 
%boundary-layer equilibrium close to the inviscid shock impingement point for all the cases. 
%An alternate criteria proposed by \citet{clauser54} fails to highlight these equilibrium regions. 
The scaling analysis for the interaction length proposed by \citet{souverein13} for turbulent
shock/boundary-layer interactions was tested for all the flow cases considered here and it was
found to be applicable for our DNS, including the transitional interactions.
Furthermore, the separation criterion $S^*_e$,
only dependent on the Mach number and flow deflection angle, correctly classified the present interactions
as attached or close to incipient separation.

Overall, our results provide numerical evidence that a transitional interaction retains the beneficial features 
of a turbulent interaction in terms of suppression of mean separation. Therefore, for practical SBLI applications,
it seems to be reasonable to trip the boundary layer a short distance upstream of the impinging shock to remove
the separation bubble, maximizing the region with a low skin friction coefficient.
Obviously, such encouraging considerations are based on a DNS database of limited extent and further
investigations are needed to completely characterize transitional SBLI. Future efforts will be devoted
to include the effects of different tripping devices, as well as to expand the range of investigated Mach- and Reynolds numbers.

\section*{Acknowledgments}
This work has been supported by the SIR program 2014 (jACOBI project, grant RBSI14TKWU),
funded by MIUR (Ministero dell'Istruzione dell'Universit\`a e della Ricerca).
The simulations have been performed thanks to computational resources provided by the Italian
Computing center CINECA under the ISCRA initiative (grant jACOBI).

\section*{References}

\bibliographystyle{unsrtnat}

\bibliography{Reference}

\begin{thebibliography}{46}
\providecommand{\natexlab}[1]{#1}
\providecommand{\url}[1]{\texttt{#1}}
\expandafter\ifx\csname urlstyle\endcsname\relax
  \providecommand{\doi}[1]{doi: #1}\else
  \providecommand{\doi}{doi: \begingroup \urlstyle{rm}\Url}\fi

\bibitem[Giepman et~al.(2016)Giepman, Louman, Schrijer, and {v}an
  Oudheusden]{giepman16}
R.~H.~M. Giepman, R.~Louman, F.~F.~J. Schrijer, and B.~W. {v}an Oudheusden.
\newblock Experimental study into the effects of forced transition on a
  shock-wave/boundary-layer interaction.
\newblock \emph{AIAA Journal}, 54(4):\penalty0 1313--1325, 2016.

\bibitem[Souverein et~al.(2013)Souverein, Bakker, and Dupont]{souverein13}
L.~J. Souverein, P.G. Bakker, and P.~Dupont.
\newblock A scaling analysis for turbulent shock-wave/boundary-layer
  interactions.
\newblock \emph{Journal of Fluid Mechanics}, 714:\penalty0 505--535, 2013.

\bibitem[Dolling(2001)]{dolling01}
D.~S. Dolling.
\newblock Fifty years of shock-wave/boundary layer interaction research: what
  next?
\newblock \emph{{AIAA} J}, 39:\penalty0 1517--1531, 2001.

\bibitem[Dupont et~al.(2006)Dupont, Haddad, and Debi{\`e}ve]{dupont06}
P.~Dupont, C.~Haddad, and J.F. Debi{\`e}ve.
\newblock Space and time organization in a shock-induced separated boundary
  layer.
\newblock \emph{J. Fluid Mech.}, {559}:\penalty0 255--277, 2006.

\bibitem[Touber and Sandham(2009)]{touber09}
E.~Touber and N.~D. Sandham.
\newblock Large-eddy simulation of low-frequency unsteadiness in a turbulent
  shock-induced separation bubble.
\newblock \emph{Theoretical and Computational Fluid Dynamics}, 23:\penalty0
  79--107, 2009.

\bibitem[Souverein et~al.(2010)Souverein, Dupont, Debi{\`e}ve, Dussauge, {van
  Oudheusden}, and Scarano]{souverein_10_b}
L.~Souverein, P.~Dupont, J.~F. Debi{\`e}ve, J.~P. Dussauge, B.~W. {van
  Oudheusden}, and F.~Scarano.
\newblock Effect of interaction strength on unsteadiness in turbulent
  shock-wave-induced separations.
\newblock \emph{AIAA J.}, 48:\penalty0 1480--1493, 2010.

\bibitem[Clemens and Narayanaswamy(2014)]{clemens14}
N.T. Clemens and V.~Narayanaswamy.
\newblock Low-frequency unsteadiness of shock wave/turbulent boundary layer
  interactions.
\newblock \emph{Annu. Rev. Fluid Mech.}, {46}:\penalty0 469--492, 2014.

\bibitem[Adamson and Messiter(1980)]{adamson_messiter80}
T.~C. Adamson and A.~F. Messiter.
\newblock Analysis of two-dimensional interactions between shock waves and
  boundary layers.
\newblock \emph{Annu. Rev. Fluid Mech.}, 12:\penalty0 103--138, 1980.

\bibitem[Delery(1985)]{delery85}
J.M. Delery.
\newblock Shock wave/turbulent boundary layer interaction and its control.
\newblock \emph{Prog.\ Aerosp.\ Sci.}, 22:\penalty0 209--280, 1985.

\bibitem[Babinsky et~al.(2009)Babinsky, Li, and Pitt~Ford]{babinsky09}
H~Babinsky, Yl~Li, and CW~Pitt~Ford.
\newblock Microramp control of supersonic oblique shock-wave/boundary-layer
  interactions.
\newblock \emph{AIAA journal}, 47\penalty0 (3):\penalty0 668, 2009.

\bibitem[Hakkinen et~al.(1959)Hakkinen, Greber, Trilling, and
  Abarbanel]{hakkinen59}
R.~J. Hakkinen, I.~Greber, L.~Trilling, and S.~S. Abarbanel.
\newblock The interaction of an oblique shock wave with a laminar boundary
  layer.
\newblock Technical Report 2-18-59W, NASA MEMO, 1959.

\bibitem[Giepman et~al.(2015)Giepman, Schrijer, and {v}an
  Oudheusden]{giepman15}
R.~H.~M. Giepman, F.~F.~J. Schrijer, and B.W. {v}an Oudheusden.
\newblock High-resolution piv measurements of a transitional shock
  wave--boundary layer interaction.
\newblock \emph{Experiments in Fluids}, 56\penalty0 (6):\penalty0 113, 2015.

\bibitem[Degrez et~al.(1987)Degrez, Boccadoro, and Wendt]{degrez87}
G.~Degrez, C.~H. Boccadoro, and J.~F. Wendt.
\newblock The interaction of an oblique shock wave with a laminar boundary
  layer revisited. {A}n experimental and numerical study.
\newblock \emph{J.~Fluid Mech.}, 177:\penalty0 247--263, 1987.

\bibitem[Katzer(1989)]{katzer89}
E.~Katzer.
\newblock On the lengthscales of laminar shock/boundary-layer interaction.
\newblock \emph{J.~Fluid Mech.}, 206:\penalty0 477--496, 1989.

\bibitem[Yao et~al.(2007)Yao, Krishnan, Sandham, and Roberts]{yao07}
Y.~Yao, L.~Krishnan, N.~Sandham, and G.~T. Roberts.
\newblock The effect of {M}ach number on unstable disturbances in
  shock/boundary-layer interactions.
\newblock \emph{Physics of Fluids}, 19\penalty0 (5):\penalty0 054104, 2007.

\bibitem[Gadd(1957)]{gadd56}
G.~E. Gadd.
\newblock A theoretical investigation of laminar separation in supersonic flow.
\newblock \emph{Journal of the Aeronautical Sciences}, 24\penalty0
  (10):\penalty0 759--771, 1957.

\bibitem[Chapman et~al.(1957)Chapman, Kuehn, and Larson]{chapmanetal_57}
D.~R. Chapman, D.~M. Kuehn, and H.~K. Larson.
\newblock Investigation of separated flow in supersonic and subsonic streams
  with emphasys on the effect of transition.
\newblock Technical Report 3869, NACA TN, 1957.

\bibitem[Greene(1970)]{greene70}
J.~E. Greene.
\newblock Interactions between shock waves and turbulent boundary layers.
\newblock \emph{Progress in Aerospace Science}, 11:\penalty0 235--340, 1970.

\bibitem[Stollery(1975)]{stollery75}
J.~L. Stollery.
\newblock Laminar and turbulent boundary layer separation at supersonic and
  hypersonic speeds.
\newblock Technical report, AGARD, 1975.

\bibitem[Delery and Marvin(1986)]{delery86}
J.~Delery and J.~Marvin.
\newblock Shock wave boundary layer interactions.
\newblock Technical Report 280, AGARDograph, 1986.

\bibitem[Gai(1977)]{gai77}
S.~L. Gai.
\newblock Shock-wave/boundary-layer interaction with suction.
\newblock \emph{Z. Flugwiss. Weltraumforsch}, 1\penalty0 (2):\penalty0 97--101,
  1977.

\bibitem[Krogmann et~al.(1985)Krogmann, Stansewsky, and Thiede]{krogmann85}
P.~Krogmann, E.~Stansewsky, and P.~Thiede.
\newblock Effects of suction on shock/boundary-layer interaction and
  shock-induced separation.
\newblock \emph{J. Aircr.}, 22\penalty0 (1):\penalty0 37--42, 1985.

\bibitem[Holden and Babinsky(2005)]{holden05}
H.~A. Holden and H.~Babinsky.
\newblock Separated shock-boundary-layer interaction control using streamwise
  slots.
\newblock \emph{J. Aircr.}, 42\penalty0 (1):\penalty0 166--171, 2005.

\bibitem[Smith et~al.(2004)Smith, Babinsky, Fulker, and Ashill]{smith04}
A.~Smith, H.~Babinsky, J.~L. Fulker, and P.~R. Ashill.
\newblock Shock-wave/boundary-layer interaction control using streamwise slots
  in transonic flows.
\newblock \emph{J. Aircr.}, 41\penalty0 (3):\penalty0 540--546, 2004.

\bibitem[Raghunathan and Mabey(1987)]{raghu87}
S.~Raghunathan and D.~G. Mabey.
\newblock Passive shock-wave/boundary-layer control on a wall-mounted model.
\newblock \emph{AIAA J.}, 25\penalty0 (2):\penalty0 275--278, 1987.

\bibitem[Mccormick(1993)]{mccormick93}
D.~C. Mccormick.
\newblock Shock/boundary-layer interaction control with vortex generators and
  passive cavity.
\newblock \emph{AIAA J.}, 31\penalty0 (1):\penalty0 91--96, 1993.

\bibitem[Anderson et~al.(2006)Anderson, Tinapple, and Surber]{anderson06}
B.~H. Anderson, J.~Tinapple, and L.~Surber.
\newblock Optimal control of shock-wave/turbulent boundary-layer interactions
  using micro-array actuation.
\newblock \emph{In 3rd AIAA Flow Control Conference, San Francisco, California,
  5-8 June}, 2006.

\bibitem[Sandham et~al.(2014)Sandham, Sch{\"u}lein, Wagner, Willems, and
  Steelant]{sandham14}
N.D. Sandham, E.~Sch{\"u}lein, A.~Wagner, S.~Willems, and J.~Steelant.
\newblock Transitional shock-wave/boundary-layer interactions in hypersonic
  flow.
\newblock \emph{Journal of Fluid Mechanics}, 752:\penalty0 349--382, 2014.

\bibitem[Davidson and Babinsky(2014)]{davidson14}
T.~S. Davidson and H.~Babinsky.
\newblock An investigation of interactions between normal shocks and
  transitional boundary layers.
\newblock \emph{44th AIAA Fluid Dynamics Conference, Atlanta, GA}, 3334, 2014.

\bibitem[Davidson and Babinsky(2015)]{davidson15}
T.~S. Davidson and H.~Babinsky.
\newblock Transition location effects on normal shock wave/boundary layer
  interactions.
\newblock \emph{53rd AIAA Aerospace Sciences Meeting, Kissimmee, Florida},
  1975, 2015.

\bibitem[Tani(1969)]{tani69}
I.~Tani.
\newblock Boundary-layer transition.
\newblock \emph{Annual Review of Fluid Mechanics}, 1\penalty0 (1):\penalty0
  169--196, 1969.

\bibitem[Bernardini et~al.(2014)Bernardini, Pirozzoli, Orlandi, and
  Lele]{bernardini14}
M.~Bernardini, S.~Pirozzoli, P.~Orlandi, and S.~K. Lele.
\newblock Parameterization of boundary-layer transition induced by isolated
  roughness elements.
\newblock \emph{AIAA Journal}, 52(10):\penalty0 2261--2269, 2014.

\bibitem[White(1974)]{white74}
F.~M. White.
\newblock In \emph{Viscous Fluid Flow}. McGraw-Hill, New York, 1974.

\bibitem[Pirozzoli et~al.(2010)Pirozzoli, Bernardini, and Grasso]{pirozzoli10}
S.~Pirozzoli, M.~Bernardini, and F.~Grasso.
\newblock Direct numerical simulation of transonic shock/boundary layer
  interaction under conditions of incipient separation.
\newblock \emph{Journal of Fluid Mechanics}, 657:\penalty0 361--393, 2010.

\bibitem[Bernardini et~al.(2016{\natexlab{a}})Bernardini, Asproulias, Larsson,
  Pirozzoli, and Grasso]{bernardini16h}
M.~Bernardini, I.~Asproulias, J.~Larsson, S.~Pirozzoli, and F.~Grasso.
\newblock Heat transfer and wall temperature effects in shock wave turbulent
  boundary layer interactions.
\newblock \emph{Phys. Rev. Fluids}, 1:\penalty0 084403, 2016{\natexlab{a}}.

\bibitem[Ducros et~al.(1999)Ducros, Ferrand, Nicoud, Darracq, Gacherieu, and
  Poinsot]{ducros99}
F.~Ducros, V.~Ferrand, F.~Nicoud, D.~Darracq, C.~Gacherieu, and T.~Poinsot.
\newblock Large-eddy simulation of the shock/turbulence interaction.
\newblock \emph{J. Comput. Phys.}, 152\penalty0 (2):\penalty0 517--549, 1999.

\bibitem[Reiss and Sesterhenn(2014)]{reiss14}
Julius Reiss and J{\"o}rn Sesterhenn.
\newblock A conservative, skew-symmetric finite difference scheme for the
  compressible navier--stokes equations.
\newblock \emph{Computers \& Fluids}, 101:\penalty0 208--219, 2014.

\bibitem[Kennedy and Gruber(2008)]{kennedy08}
C.~A. Kennedy and A.~Gruber.
\newblock Reduced aliasing formulations of the convective terms within the
  navier-stokes equations.
\newblock \emph{J. Comput. Phys.}, 227:\penalty0 1676, 2008.

\bibitem[Bernardini and Pirozzoli(2009)]{bernardini09}
M.~Bernardini and S.~Pirozzoli.
\newblock A general strategy for the optimization of runge-kutta schemes for
  wave propagation phenomena.
\newblock \emph{J. Comput. Phys.}, 228:\penalty0 4182, 2009.

\bibitem[de~Tullio et~al.(2007)de~Tullio, De~Palma, Iaccarino, Pascazio, and
  Napolitano]{detullio_07}
M.D. de~Tullio, P.~De~Palma, G.~Iaccarino, G.~Pascazio, and M.~Napolitano.
\newblock An immersed boundary method for compressible flows using local grid
  refinement.
\newblock \emph{J. Comput. Phys.}, 225\penalty0 (2):\penalty0 2098--2117, 2007.

\bibitem[Bernardini et~al.(2016{\natexlab{b}})Bernardini, Modesti, and
  Pirozzoli]{bernardini16}
Matteo Bernardini, Davide Modesti, and Sergio Pirozzoli.
\newblock On the suitability of the immersed boundary method for the simulation
  of high-reynolds-number separated turbulent flows.
\newblock \emph{Computers \& Fluids}, 130:\penalty0 84--93, 2016{\natexlab{b}}.

\bibitem[Bernardini et~al.(2012)Bernardini, Pirozzoli, and
  Orlandi]{bernardini12}
M.~Bernardini, S.~Pirozzoli, and P.~Orlandi.
\newblock Compressibility effects on roughness-induced boundary layer
  transition.
\newblock \emph{International Journal of Heat and Fluid Flow}, 35:\penalty0 45
  -- 51, 2012.

\bibitem[Acarlar and Smith(1987)]{acarlar87}
M.S. Acarlar and C.~R. Smith.
\newblock A study of hairpin vortices in a laminar boundary layer. part 1.
  hairpin vortices generated by a hemisphere protuberance.
\newblock \emph{Journal of Fluid Mechanics}, 175:\penalty0 1--41, 1987.

\bibitem[Redford et~al.(2010)Redford, Sandham, and Roberts.]{redford10}
J.~A. Redford, N.~D. Sandham, and G.~T. Roberts.
\newblock Compressibility effects on boundary-layer transition induced by an
  isolated roughness element.
\newblock \emph{AIAA Journal}, 48\penalty0 (12):\penalty0 2818 -- 2830, 2010.

\bibitem[Smits and Dussauge(2006)]{smits06}
A.~J. Smits and J.~Dussauge.
\newblock In \emph{Turbulent Shear Layers in Supersonic Flow}. Springer-Verlag
  New York, 2006.

\bibitem[Simpson(1989)]{simpson89}
R.~L. Simpson.
\newblock Turbulent boundary-layer separation.
\newblock \emph{Annual Review of Fluid Mechanics}, 21\penalty0 (1):\penalty0
  205--232, 1989.

\end{thebibliography}

\end{document}